\documentclass[twocolumn,aps,pra,showpacs,groupedaddress]{revtex4}
\usepackage{epsfig,amssymb,amsmath,mathrsfs,color} 
\usepackage[hypertex,linkcolor=red]{hyperref}
\usepackage{epsfig}

\newtheorem{lemma}{Lemma}

\newtheorem{corollary}{Corollary}
\newtheorem{theorem}{Theorem}

\newtheorem{definition}{Definition}
\def\Proof{\medskip\par\noindent{\bf Proof. }}

\input xy
\xyoption {all}
\def\>{\rangle}
\def\C{\frak{C}}
\def\rank#1{\operatorname{rank}(#1)}
\def\<{\langle}
\def\map#1{{\mathscr{#1}}}

\def\Tr{\operatorname{Tr}}
\def\kk{\rangle\!\rangle}
\def\bb{\langle\!\langle}
 
\DeclareMathOperator*{\bigast}{{\mbox{\huge$\ast$}}}
\def\biglink#1{\bigast_{#1}}
\def\combz#1#2{\operatorname{comb}({#1}_0,\dots,{#1}_{2 {#2} -1})}
\def\combu#1#2{\operatorname{comb}({#1}_1,\dots,{#1}_{2 {#2}})}
\def\tuplez#1#2{({#1}_0,\dots,{#1}_{2 {#2} -1})}
\def\tupleu#1#2{({#1}_1,\dots,{#1}_{2 {#2}})}

\def\Proof{\medskip\par\noindent{\bf Proof. }}
\def\qed{$\,\blacksquare$\par}
\def\set#1{\mathcal #1}
\def\Lin#1{\set{L}(#1)}
\def\sH{\set H}
\def\sK{\set K}
\def\sD{\set D}
\def\eg{e.~g.}
\def\ie{i.~e.}
\begin{document}

\title{A theoretical framework for quantum networks} 
\author{Giulio Chiribella}\email{chiribella@fisicavolta.unipv.it} 
\affiliation{{\em QUIT Group}, Dipartimento di Fisica  ``A. Volta'' and INFM, via Bassi 6, 27100 Pavia, Italy}
\homepage{http://www.qubit.it}
\author{Giacomo Mauro D'Ariano}\email{dariano@unipv.it}
\affiliation{{\em QUIT Group}, Dipartimento di Fisica  ``A. Volta'' and INFM, via Bassi 6, 27100 Pavia, Italy}
\homepage{http://www.qubit.it}
\author{Paolo Perinotti}\email{perinotti@fisicavolta.unipv.it} 
\affiliation{{\em QUIT Group}, Dipartimento di Fisica  ``A. Volta'' and INFM, via Bassi 6, 27100 Pavia, Italy}
\homepage{http://www.qubit.it}
\date{\today}
\begin{abstract}
  We present a framework to treat quantum networks and all possible transformations thereof,
  including as special cases all possible manipulations of quantum states, measurements, and
  channels, such as e.g. cloning, discrimination, estimation, and tomography.  Our framework is
  based on the concepts of \emph{quantum comb}---which describes all transformations achievable by a
  given quantum network---and \emph{link product}---the operation of connecting two quantum
  networks.  Quantum networks are treated both from a constructive point of view---based on
  connections of elementary circuits---and from an axiomatic one---based on a hierarchy of
  admissible quantum maps.  In the axiomatic context a fundamental property is shown, which we call
  \emph{universality of quantum memory channels}: any admissible transformation of quantum networks
  can be realized by a suitable sequence of memory channels.  The open problem whether this property
  fails for some non-quantum theory, e.g. for no-signaling boxes, is posed.
\end{abstract}
\maketitle  

\section{Introduction}
In the last decade the general description of quantum states, measurements, and transformations in
terms of density matrices, POVMs, and channels \cite{kraus,davies,holevo}, has been widely exploited
in quantum information, with many applications in high-precision measurements, quantum cryptography,
optimal cloning, quantum communication, and many others.  The success of such general description
comes from the fact that it allows one to optimize the design of quantum devices over all
possibilities admitted by quantum mechanics, thus finding the ultimate performances in the
realization of desired tasks.  Although a quantum channel can be always thought of as the result of
a unitary interaction of the system with an environment \cite{stine, ozawa}, and a POVM as a joint
von Neumann measurement on system and environment \cite{nai}, the neat advantage of using channels
and POVMs is that they simplify optimization, by getting rid of all those details that pertain
specific realizations but are irrelevant for the final purpose.

Channels and POVMs provide an efficient description of elementary circuits that transform or measure
quantum states.  When elementary circuits are combined in a larger quantum network, however, the
variety of possible tasks one can perform grows exponentially. For example, a quantum computing
network can be used as a programmable machine, which implements different transformations on input
data depending on the quantum state of the program. In some cases the program itself can be a
quantum channel, rather than a state: during computation, for instance, the network can call a
variable channel as a subroutine, so that the overall transformation of the input data is programmed
by it.  Even more generally, the action of the network can be programmed by a sequence of variable
states and channels that are called at different times, that is, at different steps of the
computation. A similar situation arises in multiple-rounds quantum games \cite{watrous}, where the
overall outcome of the game is determined by the sequence of moves (state-preparations,
measurements, and channels) performed by different players. For example, in a two-party game Alice's
strategy can be seen as a particular quantum network in which Bob's moves act as variable
subroutines.  Of course, the subroutines corresponding to Bob's moves are in turn parts of Bob's
network, so that the whole protocol can be seen as the interlinking of two networks corresponding to
Alice's and Bob's strategies.

A quantum network can be used in a number of different ways, each way corresponding to a different
kind of transformation achievable with it, e.g. transformations from states to channels, from
channels to channels, and from sequences of states/channels to channels, as discussed above.  In
fact, if we consider networks of arbitrary size, there is an infinite number of different
transformations that we can implement. This fact suggests to find new notions that generalize those
of channels and POVMs in the case of quantum networks: apparently one would have to introduce a new
mathematical object for any possible transformation. In addition, since a quantum network can
contain random circuits performing measurements and quantum operations, for any transformation one
would have to take into account also its probabilistic version.  Clearly, defining a new kind of
quantum map for any possible use of a network is not a viable approach.  On the other hand, using
the current framework based solely on states as inputs and outputs to describe a quantum network one
is presently forced to specify all elementary channels and measuring devices in it, and if one needs
to optimize the network for some desired task, then one has to face cumbersome optimization of all
its elements.  Optimizing a quantum network without suitable tools is indeed comparable to treating
tasks such as quantum error correction and quantum state estimation without the notions of channel
and POVM.

Luckily enough, an efficient treatment of quantum networks is possible, despite the infinity of
different transformations associated to them.  In this paper we will provide a complete toolbox for
the description and the optimization of quantum networks by answering the following two questions:
\emph{i)} Which are the possible tasks that a given network can accomplish?, and \emph{ii)} Which
are the transformations that a given network can undergo?  Both questions will be tackled in
Sections \ref{sec:constr} and \ref{sec:axio} from two different, complementary points of view. On
the one hand, in Section \ref{sec:constr} we will consider quantum states, POVMs, and channels, as
elementary building blocks to construct quantum networks.  The main focus will be the description of
actual networks by means of Choi-Jamio\l kowski operators, and the description of connections among
networks by means of a suitably defined composition of Choi-Jamio\l kowski operators. On the other hand,
in Section \ref{sec:axio} we will derive quantum networks and their transformations on a purely
axiomatic basis, by defining a hierarchy of admissible quantum maps.  The physical realizability of
these general transformations is proved, in a way that is similar to the unitary realization of
quantum channels: we will prove that any deterministic admissible map can be physically obtained by
a suitable sequence of memory channels.  We call this property \emph{universality of memory
  channels}, as it implies that, under mild assumptions, any deterministic transformation that is
conceivable in quantum mechanics can be always realized by some sequence of memory channels. The
case of probabilistic transformations is also considered, showing that any probabilistic
transformation can be realized by a sequence of memory channels followed by a von Neumann
measurement on some output subsystem.

\section{Preliminaries and notation}\label{sec:prel}
In this Section we list a set of elementary facts about linear maps and Choi-Jamio\l kowski
operators. The product of Choi-Jamio\l kowski operators induced by the composition of the
corresponding linear maps is defined and analyzed.
\subsection{Linear operators and linear maps}

In the following we denote with $\Lin {\sH}$ the set of linear operators on the finite dimensional
Hilbert space $\sH$. The set of linear operators from $\sH_0$ to $\sH_1$ is denoted by $\Lin{\sH_0,
  \sH_1}$. Operators $X$ in $\Lin{\sH_0,\sH_1}$ are in one-to-one correspondence with vectors $|X
\kk$ in $\sH_1 \otimes \sH_0$ as follows
\begin{equation}
\begin{split}
|X\kk &= (X \otimes I_{\sH_0}) | I_{\sH_0}\kk\\
 & = (I_{\sH_1} \otimes X^T)  |I_{\sH_1}\kk,
\end{split}\label{kk}
\end{equation}
where $I_{\sH}$ is the identity operator in $\sH$, $|I_{\sH}\kk
\in \sH^{\otimes 2}$ is the maximally entangled vector $|I_{\sH}\kk =
\sum_{n} |n\>|n\>$ (with $\{|n\>\}$ a fixed orthonormal basis for
$\sH$), and $X^T \in \Lin{\sH_1, \sH_0}$ is the transpose of $A$ with respect to the two fixed bases chosen in $\sH_0$ and $\sH_1$.

The set of linear maps from $\Lin {\sH_0}$ to $\Lin {\sH_1}$ is denoted by
$\Lin{\Lin{\sH_0},\Lin{\sH_1}}$. Linear maps $\map M$ in $\Lin{\Lin{\sH_0}, \Lin{\sH_1}}$ are in one
to one correspondence with linear operators on $\Lin{\sH_1\otimes\sH_0}$ as follows
\begin{equation}
M=\C(\map M):=\map M\otimes\map I_{\Lin{\sH_0}}(|I_{\sH_0}\kk\bb I_{\sH_0}|), 
\label{choicorr}
\end{equation}
where $\map I_{\Lin{\sH_0}}$ is the identity map on $\Lin{\sH_0}$. The
inverse map $\C^{-1}$ transforms $M\in\Lin{\sH_1\otimes\sH_0}$ into a
map in $\Lin{\Lin{\sH_0}, \Lin{\sH_1}}$ that acts on an operator $X
\in\Lin{\sH_0}$ as follows
\begin{equation}
  [\C^{-1}(M)](X)=\Tr_{\sH_0}[(I_{\sH_1}\otimes X^T)M], \label{choicorrInv} 
\end{equation}
$\Tr_{\sH}$ denoting the partial trace over $\sH$.

{\definition[Choi-Jamio\l kowski isomorphism.]\label{defchoi} The bijective correspondence $\C:\map
  M\to M$ defined through Eq.~\eqref{choicorr} is called Choi-Jamio\l kowski isomorphism. Its
  inverse $\C^{-1}:M\to\map M$ is defined through Eq.~\eqref{choicorrInv}.}

For conciseness, we will use the notation $M$ for $\C(\map M)$ throughout the paper. The operator
$M$ corresponding to the map $\map M$ is called \emph{Choi-Jamio\l kowski operator of $\map M$}.

{\lemma \label{TPchoi} A linear map $\map M$ is trace-preserving if
  and only if its Choi-Jamio\l kowski operator enjoys the following property
\begin{equation}
  \Tr_{\sH_1}[M]=I_{\sH_0}.
\label{normchan}
\end{equation}}

\Proof The trace-preserving condition writes $\Tr[\map M(X)]=\Tr[X]$.
Since
\begin{equation}
  \Tr[\map M(X)]=\Tr[(I_{\sH_1}\otimes
  X^T)M]=\Tr_{\sH_0}[X^T\Tr_{\sH_1}[M]],
\end{equation}
and $\Tr[X]=\Tr[X^T]$, the condition is satisfied for arbitrary $X$ if
and only if $\Tr_{\sH_1}[M]=I_{\sH_0}$. \qed

{\lemma \label{Hchoi} A linear map $\map M$ is Hermitian preserving if and only if
  its Choi-Jamio\l kowski operator $M$ is Hermitian.}  \Proof A map $\map M$ is
Hermitian preserving if $\map M (H)^\dag = \map M (H)$ for any
Hermitian operator $H$, or equivalently, if $\map M(X)^\dag = \map
M(X^\dag)$ for any operator $X$. The adjoint of $\map M(X)$ is
expressed as
\begin{equation}
  \map M(X)^\dag=\Tr_{\sH_0}[(I_{\sH_1}\otimes X^*)M^\dag]=\Tr_{\sH_0}[(I_{\sH_1}\otimes X^{\dag T})M^\dag].
\end{equation}
Clearly, if $M^\dag=M$ one has $\map M(X)^\dag=\map M(X^\dag)$. On the
other hand, if
\begin{equation}
  \Tr_{\sH_0}[(I_{\sH_1}\otimes X^{\dag T})M^\dag]=\Tr_{\sH_0}[(I_{\sH_1}\otimes X^{\dag T})M]
\end{equation}
for all $X$, then $M^\dag=M$, due to the Choi-Jamio\l kowski isomorphism.\qed

{\lemma \label{CPchoi} A linear map $\map M$ is completely positive (CP) if and only
  if its Choi-Jamio\l kowski operator $M$ is positive semidefinite.}  
\Proof Clearly, if $\map M$ is CP, by Eq.~\eqref{choicorr} $M\geq0$.
On the other hand, if $M\geq0$, it can be diagonalized as follows
\begin{equation}
  M=\sum_j|K_j\kk\bb K_j|,
\end{equation}
and consequently, exploiting Eqs.~\eqref{choicorrInv} and \eqref{kk},
we can write its action in the Kraus form \cite{kraus}
\begin{equation}
\map M(X)=\sum_j K_j X K_j^\dag.
\end{equation}
The Kraus form coming from diagonalization of $M$ is called {\em
  canonical}. On the other hand, since the same reasoning holds for
any decomposition $M=\sum_k|F_k\kk\bb F_k|$, there exist infinitely
many possible Kraus forms. The Kraus form implies complete positivity:
indeed, the extended map $\map M\otimes \map I_{\Lin{\sH_A}}$
transforms any positive operator $P \in \Lin{\sH_0\otimes\sH_A}$ into
a positive operator, as follows
\begin{equation}
  \map M\otimes\map I_{\Lin{\sH_A}}(P)=\sum_j (K_j\otimes I_{\sH_A}) P(K_j^\dag\otimes I_{\sH_A})\geq0.
\end{equation}
\qed

\subsection{The link product}
The Choi-Jamio\l kowski isomorphism poses the natural question on how the composition of linear maps
is translated to a corresponding composition between the respective Choi-Jamio\l kowski operators.

Consider two linear maps $\map M \in\Lin{\Lin{\sH_0},\Lin{\sH_1}}$ and $\map N \in
\Lin{\Lin{\sH_1}, \Lin{\sH_2}}$ with Choi-Jamio\l kowski operators $M \in \Lin{\sH_1 \otimes \sH_0}$
and $N\in \Lin{\sH_2 \otimes \sH_1}$, respectively. The two maps are composed to give the linear map
$\map C= \map N \circ \map M \in\Lin{\Lin{\sH_0},\Lin{\sH_2}}$. This can be easily obtained upon
considering the action of $\map C$ on an operator $X\in\Lin{\sH_0}$ written in terms of the
Choi-Jamio\l kowski operators of the composing maps
\begin{equation}
\begin{split}
  \map C&(X)=\Tr_{\sH_1}[(I_{\sH_2}\otimes\Tr_{\sH_0}[(I_{\sH_1}\otimes X^T)M]^T)N]\\
  &= \Tr_{\sH_1,\sH_0}[(I_{\sH_2}\otimes
  I_{\sH_1}\otimes X^T)(I_{\sH_2}\otimes M^{T_1})(N\otimes
  I_{\sH_0})].
\end{split}
\end{equation}
Upon comparing the above identity with the Eq. (\ref{choicorrInv}) for the map $\map C$, namely
$\map C (X) =\Tr_{\sH_0} [(I_{\sH_2} \otimes X^{T}) C]$, one obtains
\begin{equation}
C= \Tr_{\sH_1}[(I_{\sH_2}\otimes M^{T_1})(N\otimes I_{\sH_0})],
\end{equation}
where $M^{T_i}$ denotes the partial transpose of $M$ on the space $\sH_i$. 
The above result can be expressed in a compendious way by introducing the notation
\begin{equation}
N*M := \Tr_{\sH_1}[(I_{\sH_2}\otimes M^{T_1})(N\otimes
  I_{\sH_0})],
\end{equation}
which we call \emph{link product} of the operators $M \in \Lin{\sH_1
  \otimes \sH_0}$ and $N\in \Lin{\sH_2\otimes\sH_1}$. The above result
can be synthesized in the following statement.

{\theorem[Composition rules]\label{compochoi} Consider two linear maps
  $\map M \in\Lin{\Lin{\sH_0},\Lin{\sH_1}}$ and $\map N \in
  \Lin{\Lin{\sH_1}, \Lin{\sH_2}}$ with Choi-Jamio\l kowski operators
  $M \in \Lin{\sH_1 \otimes \sH_0}$ and $N\in \Lin{\sH_2 \otimes
    \sH_1}$, respectively. Then, the Choi-Jamio\l kowski operator $M
  \in \Lin{\sH_2 \otimes \sH_0}$ of the composition $\map C = \map N
  \circ \map M\in \Lin{\Lin{\sH_0},\Lin{\sH_2}}$ is given by the link
  product of the Choi-Jamio\l kowski operators $C= N*M$.}

In the following we will consider more generally maps with input and
output spaces that are tensor products of Hilbert spaces, and which
will be composed only through some of these spaces, \eg for quantum
circuits which are composed only through some wires. For describing
these compositions of maps we will need a more general definition of
link product. For such purpose, consider now a couple of operators $M
\in \Lin{\bigotimes_{m\in\set M}\sH_m}$ and
$N\in\Lin{\bigotimes_{n\in\set N} \sH_n}$, where $\set M$ and $\set
N$ describe set of indices for the Hilbert spaces, which generally
have nonempty intersection \cite{saturated-index}.

The general definition of link product then reads:

{\definition[General link product] The \emph{link product} of two
  operators $M \in \Lin {\bigotimes_{m \in\set M} \sH_m }$ and $N \in
  \Lin{\bigotimes_{n \in\set N} \sH_n}$ is the operator $M*N \in
  \Lin{ \sH_{\set N\backslash \set M} \otimes \sH_{\set M \backslash
      \set N}}$ given by
  \begin{equation}
    N*M:=\Tr_{\set M\cap \set N}[(I_{\set N\backslash \set M}\otimes M^{T_{\set M\cap \set N}})(N\otimes
    I_{\set M\backslash \set N})], 
  \end{equation}
where the set-subscript $\set X$ is a shorthand for $\bigotimes_{i\in\set X}\sH_i $, and $\set
A\backslash \set B :=\{i \in \set A, i \not \in \set B\}$ for two sets $\set A$ and $\set B$.} 

\medskip

\emph{Examples.}  For $\set M\cap\set N=\emptyset$, \eg for two operators $M$ and $N$ acting on
different Hilbert spaces $\sH_1$ and $\sH_0$, respectively, their link product is the tensor product:
\begin{equation}\label{tensorLink}
N*M=N\otimes M\in \Lin{\sH_1 \otimes \sH_0}.
\end{equation}  
For $\set N=\set M$, \ie when the two operators $M$ and $N$ act on the same Hilbert space,
the link product becomes the trace
\begin{equation}\label{traceLink}
  A* B = \Tr[A^T B].
\end{equation}

{\theorem[Properties of the link product]\label{linkprop} The
  operation of link product has the following properties:
  \begin{enumerate}
  \item $M*N=E(N*M)E$, where $E$ is the unitary swap on $\sH_{\set N\backslash \set M}
    \otimes \sH_{\set M \backslash \set N}$.  
  \item If $M_1, M_2, M_3$ act on Hilbert spaces labeled by the sets
    $\set I_1,\set I_2,\set I_3$, respectively, and $\set I_1 \cap\set I_2 \cap\set I_3=
    \emptyset $, then $M_1 *(M_2 * M_3) = (M_1 * M_2) * M_3$.
  \item If $M$ and $N$ are Hermitian, then $M*N$ is Hermitian.
  \item If $M$ and $N$ are positive semidefinite, then $M*N$ is
    positive semidefinite.
  \end{enumerate}}

\Proof Properties 1, 2, and 3 are immediate from the definition. For
property 4, consider the two maps $\map M \in
\Lin{\Lin{\sH_{\set M\backslash \set N}}, \Lin{\sH_{\set M\cap \set N}}}$ and $\map N \in
\Lin{\Lin{\sH_{\set M\cap \set N}},\Lin{\sH_{\set N\backslash \set M}}}$, associated to $M$ and
$N$ by equation Eq. (\ref{choicorrInv}). Due to Lemma \ref{CPchoi},
the maps $\map M, \map N$ are both CP. Moreover, due to Theorem
\ref{compochoi} the link product $C=N*M$ is the Choi-Jamio\l kowski operator of the
composition $\map C = \map N \circ \map M$. Since the composition of
two CP maps is CP, the Choi-Jamio\l kowski operator $C= N*M$ must be positive
semidefinite. \qed

{\bf Remark.} As it should be clear to the reader, the advantage in
using multipartite operators instead of maps is that we can associate
many different kinds of maps to the same operator $M \in \Lin
{\bigotimes_{i \in I} \sH_i}$, depending on how we group the Hilbert
spaces in the tensor product. Indeed, any partition of the set $I$
into two disjoint sets $I_0$ and $I_1$ defines a different linear map
from $\Lin {\bigotimes_{i\in I_0} \sH_i}$ to $\Lin {\bigotimes_{i\in
    I_1} \sH_i}$ via Eq. (\ref{choicorrInv}). We will see in the next
section that dealing with operators and link products allows one to
efficiently treat all possible maps associated to quantum networks.

\section{Quantum networks: constructive approach}\label{sec:constr}

\subsection{Channels and states: deterministic Choi-Jamio\l kowski operators}

In the general description of quantum mechanics, quantum states are
density matrices on Hilbert space $\sH$ of the system, i.e. positive
semidefinite operators $\rho \in \Lin{\sH}$ with $\Tr [\rho]=1$.
Deterministic transformations of quantum states are the so-called
quantum channels, a quantum channel $\map C$ from states on $\sH_0$ to
states on $\sH_1$ being a trace-preserving completely positive map.
According to Lemmas \ref{TPchoi}, \ref{Hchoi}, \ref{CPchoi}, the Choi-Jamio\l kowski
operator corresponding to $\map C$ is a positive semidefinite operator
$C \in \Lin {\sH_1 \otimes \sH_0}$ satisfying $\Tr_{\sH_1}[C] =
I_{\sH_0}$.

It is immediate to see that a density matrix is a particular case of
Choi-Jamio\l kowski operator of a channel, namely a Choi-Jamio\l kowski operator with
one-dimensional input space $\sH_0$: in this case the condition
$\Tr_{\sH_1}[C] = I_{\sH_0}$ becomes indeed $\Tr[C] =1$. This reflects
the fact that having a quantum state is equivalent to having at
disposal one use of a suitable preparation device. The application of
the channel $\map C$ to the state $\rho$ is equivalent to the
composition of two channels, and is indeed given by the link product
of the corresponding Choi-Jamio\l kowski operators
\begin{equation}
\map C (\rho) = C * \rho,
\end{equation}
which agrees both with Eq. (\ref{choicorrInv}) and Theorem \ref{compochoi}.

The opposite example is the completely demolishing ``trace channel'' $\map T
(\rho)= \Tr[\rho]$, which transforms quantum states into their
probabilities (of course, normalized density matrices give unit probabilities): this channel has one-dimensional output space $\sH_1$,
and, accordingly its Choi-Jamio\l kowski operator is $T = I_{\sH_{0}}$.  Notice that
the normalization of the Choi-Jamio\l kowski operator $C \in \Lin{\sH_1 \otimes
  \sH_0}$ of a channel $\map C$ can be also written in terms of
concatenation with the trace channel as
\begin{equation}\label{channelnormLink}
C * I_{\sH_1} = I_{\sH_0}. 
\end{equation}

\subsection{Instruments, random sources, and POVMs: probabilistic Choi-Jamio\l kowski operators}

In addition to the Choi-Jamio\l kowski operators of deterministic quantum devices,
one can consider their probabilistic versions.  A complete family of
probabilistic transformations from states on $\sH_0$ to states on
$\sH_1$, known as \emph{quantum instrument}, is a set of CP maps $\{\map C_i
~|~ i \in I \}$ summing up to a trace-preserving CP map $\map C =
\sum_{i\in I} \map C_i$. The corresponding Choi-Jamio\l kowski operators $\{C_i~|~ i
\in I\}$ are positive semidefinite operators summing up to a
deterministic Choi-Jamio\l kowski operator $C = \sum_{i\in I} C_i$ with $C *
I_{\sH_1} = I_{\sH_0}$. For families of probabilistic transformations,
the index $i$ has always to be intended as a classical outcome, that is known to the
experimenter, and heralds the occurrence of different random transformations.

For one-dimensional input space $\sH_0$, a complete family of
probabilistic Choi-Jamio\l kowski operators $\{\rho_i ~|~ i \in I\}$ with $\sum_i
\rho_i =\rho, \Tr [\rho]=1$ describes a \emph{random source of quantum
states}. Applying the trace channel $\map T$ after the source gives the
probability of the source emitting the $i$-th state:  $p_i = \Tr[\rho_i] = \rho_i
*I_{\sH_1} $ (of course $p_i \ge 0$ and $\sum_i p_i =1$).

For one-dimensional output space $\sH_1$, a complete family of
probabilistic Choi-Jamio\l kowski operators is instead a POVM $\{P_i ~|~ i \in I \}$,
$\sum_i P_i = I_{\sH_1}$.  Measuring the POVM on the state $\rho$ is
equivalent to applying the random device described by $\{P_i\}$ after
the preparation device for state $\rho$, producing as the outcome the
probabilities
\begin{equation}\label{bornruleLink}
p(i|\rho) = \rho * P_i = \Tr[ \rho P_i^T].
\end{equation} 
Apart from the transpose, which can be absorbed in the definition of
the POVM, this is nothing but the Born rule for probabilities,
obtained here from the composition of a preparation channel with a
random transformation with one-dimensional output space.

In conclusion, states, channels, random sources, instruments, and
POVMs can be treated on the same footing as deterministic and
probabilistic transformations, which in turn can be described using
only Choi-Jamio\l kowski operators and link product.

\subsection{Quantum networks and memory channels}

In the previous Subsections we have shown that all elementary quantum
circuits can be described in terms of Choi-Jamio\l kowski operators and their link
products. Here this approach is exploited to describe quantum
networks, as a result of the composition of such elementary circuits.
This is the approach outlined in Ref \cite{qca}.

\subsubsection{Topology, causal ordering, and sequential ordering}

A quantum network is obtained by assembling a number of elementary
circuits, each of them represented by its Choi-Jamio\l kowski operator. In the
remainder of the paper we adopt the following convention, which
appears to be very convenient for the description of quantum networks:
if an elementary circuit is run more than once, i.e. at different
steps of the computation, we attach to each different use a different
label, so that different uses of the same circuit are actually
considered as different circuits.

To build up a particular quantum network one needs to have at disposal
the whole list of elementary circuits and a list of instructions about
how to connect them.  In connecting circuits there are clearly two
restrictions: \emph{i)} one can only connect the output of a circuit
with the input of another circuit, and {\emph{ii)} there cannot be
  cycles \cite{cycles}.  These restrictions ensure causality, namely
  the fact that quantum information in the network flows from input to
  output without loops. This implies that the connections in the
  quantum network can be represented in a directed acyclic graph
  (DAG), where each vertex represents a quantum circuit, and each
  arrow represents a quantum system traveling from one circuit to
  another, as in Fig.  \ref{dags}a.  Notice that such a graph
  represents only the internal connections of the networks, while to
  have a complete graphical representation one should also append to
  the vertices a number of free incoming and outgoing arrows
  representing quantum systems that enter or exit the network. In
  other words, the graphical representation of a quantum network is
  provided by a DAG where some sources (vertices without incoming
  arrows) and some sinks (vertices without outgoing arrows) have been
  removed, as in Fig. \ref{dags}b.  The free arrows remaining after
  removing a source represent input systems entering the network,
  while the free arrows remaining after removing a sink represent
  output systems exiting the network.

\begin{figure}[h]
\epsfig{file=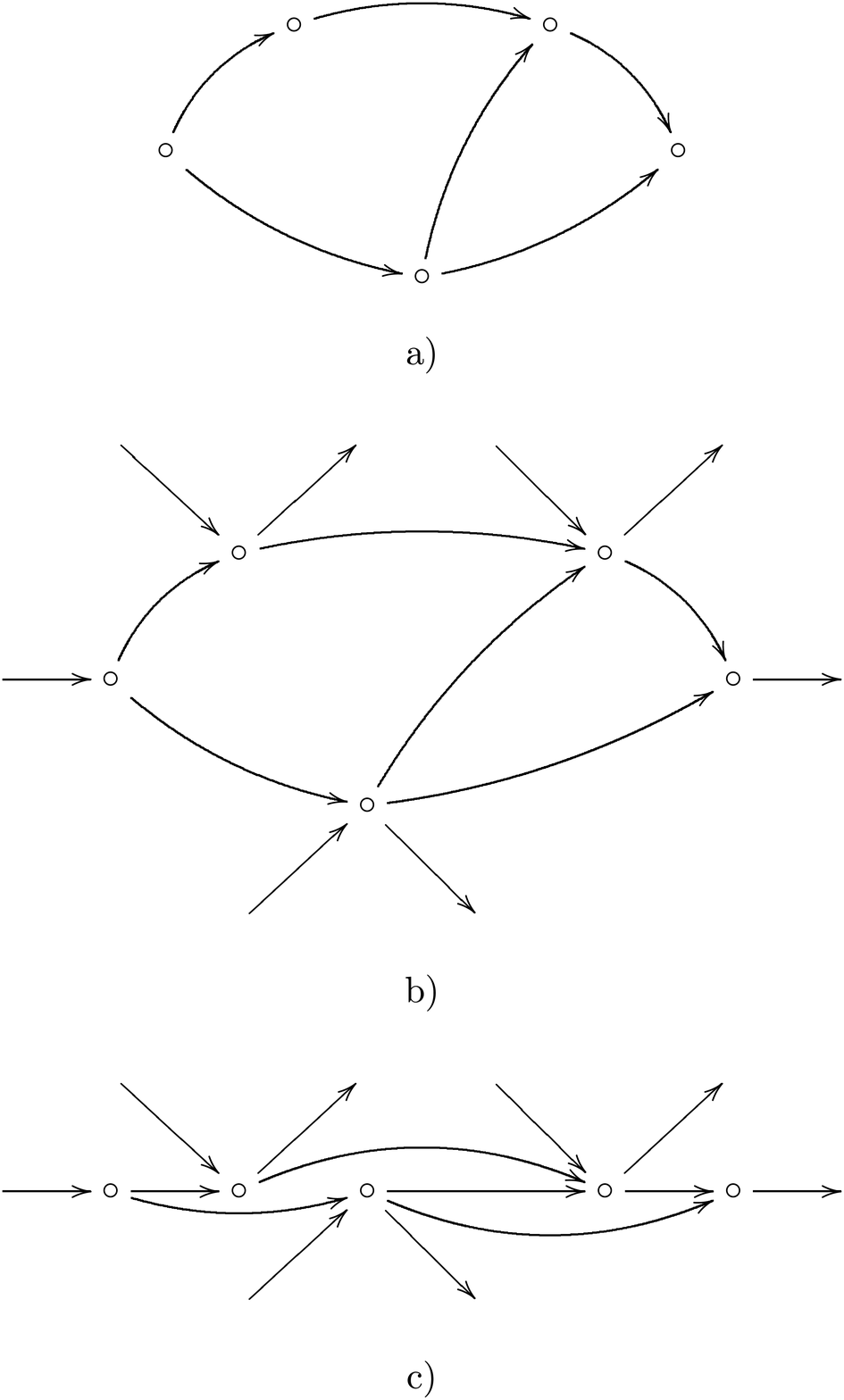,width=\columnwidth}
  \caption{\label{dags} a) Graphical representation of internal
    connections in a quantum network: vertices represent quantum
    operations, incoming and outgoing arrows represent input and
    output systems. The resulting diagram is a direct acyclic graph.
     b) Graphical representation of a quantum network: free
    incoming (outgoing) arrows have been added to the diagram in a) in order to
    represent input (output) systems entering (exiting) the network.
    c) Totally ordered quantum network. The vertices in diagram b)
    have been ordered from left to right according to a sequential
    ordering compatible with the causal ordering fixed by input/output
    relations.}
\end{figure}

The flow of quantum information along the arrows of the graph induces
a partial ordering of the vertices: we say that the circuit in vertex
$v_1$ \emph{causally precedes} the circuit in vertex $v_2$ ($v_1
\preceq v_2$) if there is a directed path from $v_1$ to $v_2$.  A well
known theorem in graph theory states that for a directed acyclic graph
there always exists a way to extend the partial ordering $\preceq$ to
a total ordering $\le$ of the vertices. Intuitively speaking, the
relation $\le$ fixes a schedule for the order in which the circuits in
the network can be run, compatibly with the causal ordering of
input-output relations. In general, the total ordering $\le$ is not
uniquely determined by the partial ordering $\preceq$: the same
quantum network can be used in different ways, corresponding to
different orders in which the elementary circuits are run.

A quantum network with a given sequential ordering of the vertices
becomes a compound quantum circuit, in which different operations are
performed according to a precise schedule. Totally ordered quantum
networks have a large number of applications in quantum
information, and, accordingly, they have been given different names,
depending on the context. For example, they are referred to as
\emph{quantum strategies} in quantum game theoretical and
cryptographic applications \cite{watrous}. Moreover, a totally ordered
quantum network is equivalent to a sequence of channels with memory,
as illustrated in Fig. \ref{netmemo} a.  Currently, the most studied
case in the literature on memory channels is that in which all
channels of the sequence are identical, as represented in Fig.
\ref{netmemo} b): here the memory must be first initialized in some
fixed state $|0\>$, and eventually traced out. Clearly, the network in
Fig. \ref{netmemo} b) is the particular case of that in Fig.
\ref{netmemo} a) corresponding to $ \map C_0 (\rho) = \map C (\rho
\otimes |0\>\<0| )$, $\map C_2 = \map C_3 = \map C_{N-2} = \map C$,
and $\map C_{N-1}(\rho) = \Tr_{M} [\map C (\rho) ]$, $\Tr_M$ being the
partial trace over the memory system.
\begin{figure}[h]
\epsfig{file=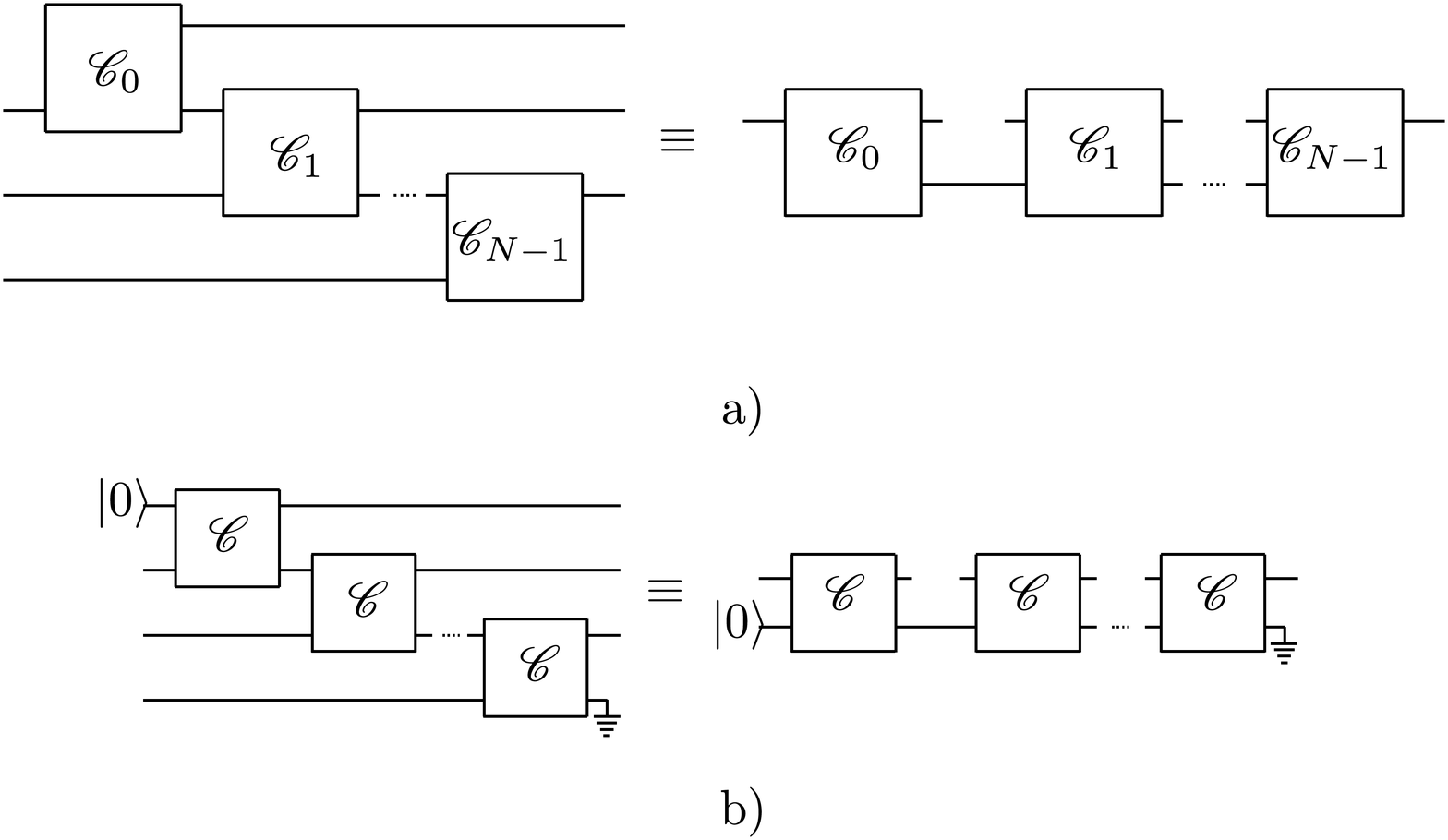,width=\columnwidth}
\caption{\label{netmemo} a) Equivalence between an arbitrary sequence
  of memory channels and a totally ordered quantum network: a sequence
  of quantum memory channels from Alice (left side) to Bob (right
  side) is equivalent modulo stretching and reshuffling of the quantum
  wires to an array of channels connected by internal ancillae, i.e. a
  totally ordered quantum network. b) A sequence consisting of
  identical memory channels, with the memory initialized in state
  $|0\>$ before the first use, and traced out after the last use. }
\end{figure}

In the following we will be always interested in quantum networks
equipped with a total ordering of the vertexes, and, accordingly, the
expressions ``quantum network'', ``quantum strategy'', and ``sequence of memory
channels'' will be used as synonymous.


\subsubsection{Deterministic quantum networks}

We start here by considering deterministic quantum networks, i.e.
networks which do not produce random transformations. A deterministic
quantum network is composed by deterministic quantum circuits, i.e.
quantum channels. Let $\{\map C_j ~|~ j \in V\}$ be the channels
corresponding to the vertices of the graph, and $\{C_j~|~j \in V \}$
their Choi-Jamio\l kowski operators.

Let us consider a network with a finite number of vertices $N = |V|
< \infty$, and let us label the vertices with numbers from 0 to $N-1$,
according to the sequential ordering of the network. The Hilbert
spaces of each Choi-Jamio\l kowski operator $C_j$ are labeled by indices in the sets
$A_j^-$ ($A_j^+$) of incoming (outgoing) arrows at vertex $j$, the
elements in $A_j^-$ ($A_j^+$) corresponding to input (output) systems
of the quantum channel $\map C_j$. Let $A_j = A_j^- \cup A_j^+$ be the
set of all arrows at vertex $j$, and let $\sH_{A_j^-}, \sH_{A_j^+},$
and $\sH_{A_j}$ be the tensor products of all Hilbert spaces
associated to the sets $A_j^-, A_j^+$, and $A_j$, respectively.  Then,
the normalization of the channel $C_j$ reads
\begin{equation}\label{normvertex}
I_{A_j^+} * C_j = I_{A_j^-},
\end{equation}  
which comes from Eq. (\ref{channelnormLink}).

Since $A_i \cap A_j \cap A_k = \emptyset$ for any $i,j,k = 0,1, \dots
,N$, we can always define the link product $C_i * C_j * C_k$ (the link
product is associative due to Theorem \ref{linkprop}). Accordingly, we
can define the \emph{Choi-Jamio\l kowski operator of the network} as
\begin{equation}
  R^{(N)}= C_0 * C_1 * \dots * C_{N-1} = \biglink{j\in V} ~  C_j.
\end{equation}
Let us denote by $\sH_{2j}$ and $\sH_{2j+1}$ the Hilbert spaces of all
free (i.e. not connected) input and output systems at vertex $j$,
respectively.  
Since the Hilbert spaces of the connected systems are traced out in
the link product, it is immediate to see that the Choi-Jamio\l kowski operator of the
network is an operator $R^{(N)}$
on 
$\bigotimes_{j=0}^{2N-1} \sH_j$.

The normalization of the Choi-Jamio\l kowski operator of the network is given by the
following condition:

{\lemma \label{lemrecnet} {\bf (Normalization condition)} Let
  $R^{(N+1)} \in \Lin {\bigotimes_{j=0}^{2N+1} \sH_j}$ be the
  Choi-Jamio\l kowski operator of a deterministic quantum network with
  $N+1$ vertices , ordered from $0$ to $N$.  Then, $R^{(N+1)}$ is positive semidefinite and
  satisfies the relation
\begin{equation}\label{rec}
I_{2N+1} * R^{(N+1)} = I_{2N} * R^{(N)},
\end{equation}
where $R^{(N)} \in \Lin {\bigotimes_{j=0}^{2N-1} \sH_j} $ is the Choi-Jamio\l kowski
operator of a network with $N$ vertices ordered from $0$ to $N-1$.}

Notice that in terms of partial traces and tensor products the
normalization of the Choi-Jamio\l kowski operator $R^{(N+1)}$ can be
equivalently written in the (less symmetric) form
\begin{equation}
  \Tr_{2N+1} \left[ R^{(N+1)}\right] = I_{2N} \otimes R^{(N)}.
\end{equation} 

\Proof Denote by $\sH_{\overline {2N}}$ the Hilbert spaces of all
incoming internal connections at vertex $N$, so that $\sH_{A_N^-} =
\sH_{2N} \otimes \sH_{\overline{2N}}$. We have $I_{2N+1} * R^{(N)} =
C_0 * \dots * C_{N-1} * (I_{2N+1} * C_N)$. Since $N$ is the last
vertex, all outgoing arrows are free, i.e.  $\sH_{A_N^+} =
\sH_{2N+1}$. Therefore the normalization of the channel $\map C_N$
(Eq.  (\ref{normvertex})) gives $I_{2N+1} * C_N = I_{A_{N}^-} = I_{2N}
\otimes I_{\overline{2N}} = I_{2N} * I_{\overline{2N}}$ (see Eq.
(\ref{tensorLink}) for the last equality).  We then obtain $I_{2N+1} *
R^{(N)} = I_{2N} * C_0 * \dots * C_{N-2} * C'_{N-1}$, where $C'_{N-1}
= I_{\overline{2N}} * C_{N-1}$ is the Choi-Jamio\l kowski operator of the channel
$\map C_{N-1}$ followed by the partial trace over the space
$\sH_{\overline{2N}}$. Clearly, $R^{(N-1)} = C_1 * \dots * C_{N-2} *
C_{N-1}'$ is the Choi-Jamio\l kowski operator of a network with $N$ vertices. \qed

Iterating the above result we then have the following: {\corollary Let
  $R^{(N)} \in \Lin{\bigotimes_{j=0}^{2N-1} \sH_j}$ be the Choi-Jamio\l kowski
  operator of a quantum network with $N$ vertices. Then, $R^{(N)} \ge
  0$ and the following relations hold
\begin{equation}
    \begin{split}
    &\Tr_{2j-1}[R^{(j)}]=I_{2j-2}\otimes R^{(j-1)},\quad 2\leq j\leq N\\
    &\Tr_1[R^{(1)}]=I_0,
  \end{split}
\end{equation}
each $R^{(j)}$ being a suitable positive operator on
$\bigotimes_{k=0}^{2j-1} \sH_k$.}

We conclude the paragraph by noting that also the converse of Lemma
\ref{lemrecnet} can be proved. The proof is essentially based on the
same argument as in Ref. \cite{KW} (uniqueness of the minimal
Stinespring dilation).

\begin{figure}[h]
\epsfig{file=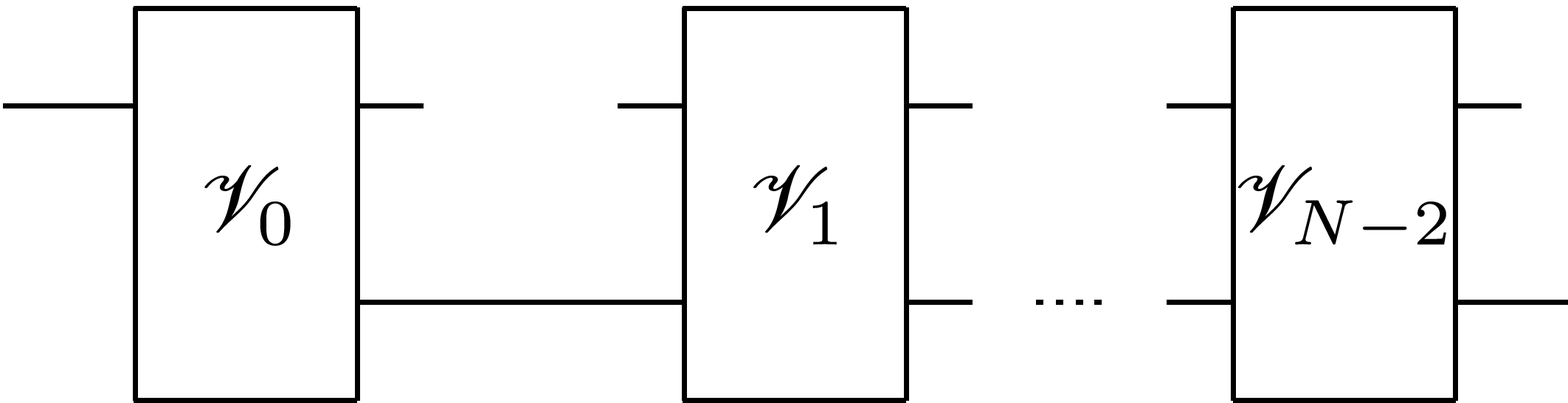,width=\columnwidth}
\caption{\label{isometries} Quantum network resulting from a
  concatenation of $N$ (generally different) isometric channels $\map
  V_j (\rho) := V_j \rho V_j^{\dag}$, with the last channel followed
  by partial trace over the ancillary degrees of freedom.  Any
  positive operator satisfying Eq. (\ref{normcondrec}) is the Choi-Jamio\l kowski
  operator of a network of this form.}
\end{figure}

{\theorem \label{realchoi} Let $R^{(N)} \in \Lin
  {\bigotimes_{j=0}^{2N-1} \sH_j}$ be a positive operator satisfying
  the relations
\begin{equation}
    \begin{split}
    &\Tr_{2j-1}[R^{(j)}]=I_{2j-2}\otimes R^{(j-1)},\quad 2\leq j\leq N\\
    &\Tr_1[R^{(1)}]=I_0.
  \end{split}
\label{normcondrec}
\end{equation}
where $R^{(j)}, 1 \le j \le n-1$ are suitable positive operators. Then
$R^{(N)}$ is the Choi-Jamio\l kowski operator of a quantum network.}  \Proof First,
notice that each $R^{(j)}$ is the Choi-Jamio\l kowski operator of a channel $\map R^{(j)}$
from states on the even Hilbert spaces $\bigotimes_{k=0}^{j-1}
\sH_{2k}$ to states on the odd Hilbert spaces $\bigotimes_{k=0}^{j-1}
\sH_{2k+1}$. Indeed, Eq. (\ref{normcondrec}) implies that
\begin{equation}
  \Tr_{1,3, \dots, 2j-1} [R^{(j)}] = I_{0, 2, \dots, 2j-2},
\end{equation} 
whence $\map R^{(j)}$ is trace preserving due to Lemma (\ref{TPchoi}).
The problem is then to show that the multipartite channel $\map
R^{(N)}$ arises from the concatenation of $N$ channels as in Fig.
\ref{isometries}. In particular, we show that $\map R^{(N)}$ can be
obtained as a concatenation of $N$ isometries. The proof is by
induction. For $N=1$ the statement is equivalent to Stinespring's
dilation of channels \cite{stine}: the Kraus operators of the channel
$\map R^{(1)}$ define an isometry $W^{(1)} = \sum_i |i\>_A \otimes
K^{(1)}_i$, where $\{|i\>_A\}$ are orthonormal states for an ancilla
$A$.  As the induction hypothesis, we suppose now that the isometry
$W^{(N)}:=\sum_i |i\>_A\otimes K^{(N)}_i$, defined by the canonical
Kraus operators of $\map R^{(N)}$, arises from the concatenation of
$N$ isometries, as in Fig.  \ref{isometries}. Using such hypothesis,
we then prove that also the isometry $W^{(N+1)} = \sum_i |i\>_B
\otimes K^{(N+1)}_i$ is the concatenation of $N+1$ isometries as in
Fig. \ref{isometries}.  Indeed, using Eq. (\ref{choicorrInv}), it is
immediate to see that the condition
\begin{equation}
  \Tr_{2N+1}[R^{(N+1)}]=I_{2N}\otimes R^{(N)},
\end{equation}
implies that
\begin{equation}
  \Tr_{2N+1} [\map R^{(N+1)} (\rho)] = \map R^{(N)} \left( \Tr_{2N}[\rho] \right),
\end{equation}
for any state $\rho$ on $\bigotimes_{j=0}^{N} \sH_{2j}$. Therefore $\{
\<m| K_i^{(N+1)}\}$ and $\{ K^{(N)}_j \otimes \<n| \}$ are two Kraus
representations of the same channel, the latter being canonical, as $\Tr
\left [K^{(N) \dag}_i K_{i'}^{(N)} \otimes |n\>\<n'| \right] =
\delta_{nn'} \delta_{ii'}$. Since any Kraus representation is
connected to the canonical one by the matrix elements of an isometry,
we have
\begin{equation}
  \<m|K^{(N+1)}_i=\sum_{nj} V_{mi,nj}~ K^{(N)}_j \otimes \<n|,
\end{equation}
or, equivalently
\begin{equation}
  K^{(N+1)}_i={}_B\<i|(V\otimes I_{2N+1,\dots, 1})(I_{2N}\otimes W^{(N)}),
\end{equation}
where $V = \sum_{mi,nj} V_{mi,nj} |m\>\<n| \otimes {}_B|i\>\<j|_A$ is
an isometry from $\sH_{2N}\otimes \sH_A$ to $\sH_{2N+1}\otimes\sH_B$.
Therefore we have
\begin{equation}
  \begin{split}
    W^{(N+1)} &= \sum_{i} |i\>_B \otimes K^{(N+1)}_i\\
    & = (V\otimes I_{2N+1,\dots, 1})(I_{2N}\otimes W^{(N)})
  \end{split}
\end{equation}
Accordingly, the map $\map R^{(N+1)}$ can be expressed as
\begin{equation}
  \map R^{(N+1)}(\rho)=\Tr_B[(V\otimes I)(I\otimes W^{(N)})\rho(I\otimes W^{(N)\dag})(V^\dag\otimes I)],
\end{equation}
where $V$ maps the $(2N)$-th system and ancilla $A$ to the $(2N+1)$-th
system and ancilla $B$. Along with the induction hypothesis, this
proves the theorem. \qed

\subsubsection{Network complexity}

In theorem \ref{realchoi} we proved that quantum networks are in one
to one correspondence with Choi-Jamio\l kowski operators satisfying
the conditions in Eq.~\eqref{normcondrec}. In particular, the proof
involves the minimal Stinespring isometry of the channel $\map
R^{(N)}$ from states on $\sH_{\mathrm{in}}:=\bigotimes_{k=0}^{j-1}
\sH_{2k}$ to states on $\sH_{\mathrm{out}}:=\bigotimes_{k=0}^{j-1}
\sH_{2k+1}$. In Ref \cite{covinst}, the expression of the minimal
Stinespring isometry in terms of the Choi-Jamio\l kowski operator was
derived
\begin{equation}
  W^{(N)}=\left(I\otimes \sqrt{R^{(N)*}}\right)|I\kk_{\mathrm{out},\mathrm{out}}\otimes I_{\mathrm{in}},
  \label{minist}
\end{equation}
Where the ancillary Hilbert space is isomorphic to a subspace of
$\sH_{\mathrm{out}}\otimes\sH_{\mathrm{in}}$ with dimension equal to
the rank of the Choi-Jamio\l kowski operator $R^{(N)}$. Repeating the
same argument for each $R^{(j)}$ with $1\leq j\leq N-1$, one obtains
an isometric extension of the channel $\map R^{(N)}$ with $N$
ancillae, each one appearing at a vertex $j-1$ and disappearing at the
subsequent vertex $j$ (apart from the last one, appearing at vertex
$N-1$ and purifying the output system).  The maximum
$d_\mathrm{max}:=\max_{1\leq j\leq N}\rank{R^{(j)}}$ denotes the
maximum dimension of the ancilla required by the network described by
$R^{(N)}$. Moreover, if one defines
$r_j:=\rank{R^{(j)}}\max\{d_{2j+1},d_{2j+2}\}$, for $0\leq j\leq N-2$,
and $r_{N-1}:=\rank{R^{(N)}}d_{2N-1}$, the number
\begin{equation}
  r(R^{(N)}):=\max_{0\leq j\leq N-1}r_j
\end{equation}
is the maximal dimension that must be coherently controlled in order
to implement the network. We can say that the quantity
$d_\mathrm{max}$ describes the complexity of the network corresponding
to the Choi operator $\map R^{(N)}$ in terms of quantum memory needed,
while $r(R^{(N)})$ describes the complexity of the network in terms of
coherent control. However, $d_\mathrm{max}$ and $r(R^{(N)})$ provide
only upper bounds on the actual memory and coherence control complexity.
Indeed, in the Stinespring isometric extension coherence of ancillary
systems is preserved up to the last step. However, it can often happen
that some ancillary subsystem interacts with the systems only at
vertex $j$. In this case, one could trace out such subsystem just
after the interaction at vertex $j$.\par

On the other hand, the analysis of complexity in terms of number of
elementary gates needed requires a detailed description of all the
unitaries that one must use to implement the isometries $W^{(j)}$.

\subsubsection{Probabilistic quantum networks}

A probabilistic quantum network is a network in which the channels
$\{\map C_n~|~ n \in V\}$ are replaced by quantum instruments $\{\map
C_{n,i_n}~|~ n\in V\}$, where $i_n$ is the label of the random
transformation taking place at vertex $n$ (in practical terms, the
outcome of the $n$-th measurement). Defining the set $I$ of
poly-indices $i =(i_0,i_1, \dots, i_N)$ corresponding to measurement
outcomes, we have a family $\{R^{(N)}_{i}\}$ of Choi-Jamio\l kowski operators of the
probabilistic network, given by
\begin{equation}
  R^{(N)}_{i} = C_{0,i_0} * \dots * C_{N,i_N},\ i\in I
\end{equation}
Clearly, the sum of the operators $R_i^{(N)}$ over $i$ gives the Choi-Jamio\l kowski
operator of a deterministic quantum network. Moreover, also the
converse statement is true:

{\theorem\label{probnetchoi} Let $\{R_i^{(N)} \in
  \Lin{\bigotimes_{j=0}^{2N-1} \sH_j}~|~i = 1, \dots, k ~\}$ be a
  collection of operators with the properties \emph{i)} $R_i^{(N)}\ge
  0$ and $\sum_{i=1}^k R_i^{(N)} = R^{(N)}$, with $R^{(N)}$
  satisfying the relations of Eq. (\ref{rec}). Then, each $R_i^{(N)}$
  is the Choi-Jamio\l kowski operator of the probabilistic quantum network,
  consisting of $N$ isometric interactions, followed by a von Neumann
  measurement on a $k$-dimensional ancilla giving outcome $i$, as in
  Fig. \ref{fig:probnet}.}

\begin{figure}[h]
\epsfig{file=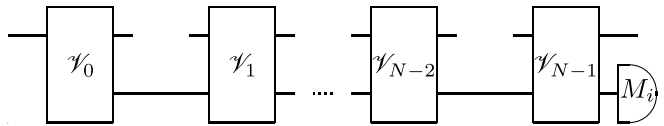,width=\columnwidth}
\caption{\label{fig:probnet} Quantum network resulting from a
  concatenation of $N$ isometric channels, with the last channel
  followed by a von Neumann measurement $\{M_i\}$ over the ancillary
  degrees of freedom.  Any collection of positive operators
  $\{R_i^{(N)}\}$ summing up the the Choi-Jamio\l kowski operator of a deterministic
  quantum network describes a probabilistic quantum network of this
  form.}
\end{figure}

\Proof Let us consider the following Choi-Jamio\l kowski operator
\begin{equation}
 \tilde  R^{(N)}:=\sum_{i\in \set I} R^{(N)}_i\otimes|i\>\<i|_A,
\end{equation}
where $|i\>_A, i=1, \dots, k$ is an orthonormal basis for an ancillary
Hilbert space $\sH_A$.  Using Eq. (\ref{rec}) and Theorem
\ref{realchoi}, it is immediate to see that $R^{(N)}$ is the
Choi-Jamio\l kowski operator of a deterministic quantum network with
$N$ vertices, the last vertex having the output space $\tilde
\sH_{2N-1}:=\sH_{2N-1} \otimes \sH_A$. In particular, we know that
$R^{(N)}$ can be realized by a sequence of isometric channels. Now
apply the von Neumann measurement given by $\{M_i= |i\>\<i|\}$ on the
ancilla $\sH_A$.  Conditionally to outcome $i$, the Choi-Jamio\l
kowski operator of the network will be $\tilde R^{(N)} * M_i = \<i |
\tilde R^{(N)} |i\>_A = R^{(N)}_i$, where we used Eq. (\ref{traceLink}) .
\qed

\subsubsection{Transformations achievable with a given quantum
  network}

Given a network of quantum circuits, we can perform a number of
different tasks. We can use the network as a programmable device, by
feeding into it some quantum systems acting as the program, or we can
connect some outputs with some inputs through a set of external
circuits. Alternatively, we can make measurements on some outputs and
decide accordingly which states to send to the next inputs, or we
simply can use the network as a single multipartite channel.  Any
different use of a quantum network, however, will be always equivalent
to the connection of the network with another quantum network, as in
Fig.  \ref{netconnect}.  

Connecting two networks with vertices $V$ and $W$, respectively, means
composing the corresponding graphs by joining some of the free
outgoing arrows of a network with the free incoming arrows of the
other, in such a way that the new graph is still a directed acyclic
graph, with vertices $U = V \cup W$. Again, the requirement that the
graph of connections in the composite network is acyclic is crucial in
order to have a network where quantum information flows from input to
outputs without loops.  We adopt the convention that if two
vertices $v\in V$ and $w \in W$ are connected by joining two arrows,
the two quantum systems corresponding to such arrows are identified
(see Fig.  \ref{netconnect}).

\begin{figure}[h]
\epsfig{file=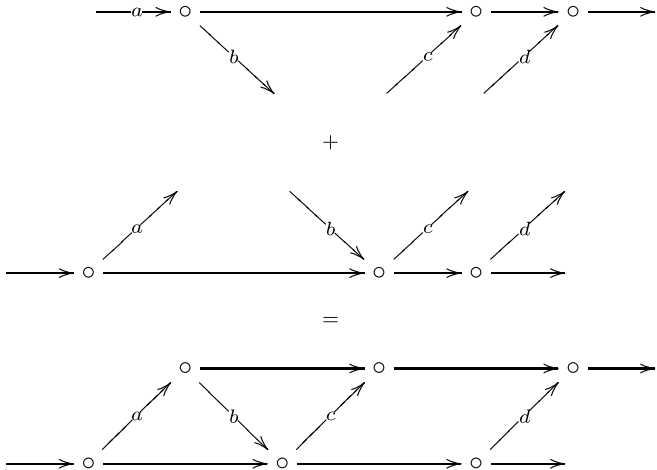,width=\columnwidth}
  \caption{\label{netconnect} The scheme represents the connection of
    two networks, in which junction of two arrows means identification
    of the corresponding quantum systems. The final network is still a
    direct acyclic graph, with the set of vertices coinciding with the
    union of sets of vertices of the component sub-networks, and with
    some free input and output arrows.}
\end{figure}

Let us proceed to determine the Choi-Jamio\l kowski operator resulting from the
composition of two networks, with $|V|=N$ and $|W| =M$ vertices
respectively, and with a given ordering of the vertices.  Notice that,
although the order of vertices within each network is fixed, a priori
there is no relative ordering between vertices of one network and
vertices in the other. However, once we fix a legitimate way of
connecting the two networks we can also define a total ordering of the
vertices which is compatible with the causal flow of quantum
information in the composite network. In other words, we can order the
vertices $U=V \cup W$ of the composite network by labeling them with
numbers from $0$ to $N+M-1$. With this labeling, $V$ and $W$ become
two disjoint partitions of the set $\{0, 1, \dots, N+M-1\}$.  We then
have the following

{\corollary\label{theo:jointwonet} Let $R$ and $S$ be the Choi-Jamio\l kowski operators of two
  quantum networks. The Choi-Jamio\l kowski operator of the network resulting from their composition
  (output of $R$ fed into the input of $S$) is given by 
\begin{equation}\label{twonet}
  T = S* R
\end{equation}}

\Proof The proof is an immediate consequence of associativity of the link product.
\qed

A possible way of transforming a given network is to connect it with
another network containing state preparations and measurements, so
that the resulting network has neither incoming nor outgoing
quantum systems.  In this case, any measurement outcome corresponds to
a probabilistic transformation, which turns the input network into a
probability.  Corollary \ref{theo:jointwonet} shows that the probabilities in such an
experiment will be given by the generalized Born rule

\begin{equation}\label{genborn}
  p(i|R) = R * S_i = \Tr [R S^{T}_i].
\end{equation}
This means that two networks with the same Choi-Jamio\l kowski operator $R$ are experimentally
indistinguishable. More precisely, as long as one is not interested in the internal functioning of the
network and is only concerned only with its input/output relations, two networks with the same
Choi-Jamio\l kowski operator are indistinguishable. 

In conclusion, the action of a quantum network can be completely identified by
its Choi-Jamio\l kowski operator.  Notice that, moreover, the Choi-Jamio\l kowski operator provides
a much simpler description of a quantum network than the list of all channels and all connections
among them. Indeed, the operator $R$ acts only on the Hilbert spaces of the quantum systems that
actually enter and exit the network, and \emph{not} of the quantum systems that are internal to the
network. 

As we will see in the following section, the Choi-Jamio\l kowski operator of a quantum network
coincides with the quantum comb, an abstract object that can be derived on a purely axiomatic basis.

\section{Axiomatic approach: the hierarchy of admissible quantum
  maps}\label{sec:axio}

While in the previous Section we focused on the description of
transformations that can be achieved by assembling elementary circuits
into networks and by connecting networks with each other, in the
following we take an axiomatic point of view, aimed to classify the
transformations that are admissible \emph{in principle} according to
quantum mechanics.  With ``admissible transformations'' we mean here
general input/output transformations that \emph{i)} are compatible
with the probabilistic structure of the theory, and \emph{ii)} produce
a legitimate output when applied locally on one side of a bipartite
input.  Such transformations are defined recursively, by starting from
channels and quantum operations, and progressively generating an
infinite hierarchy of quantum maps.  Despite the hierarchy of
transformations is unbounded, we will show that a dramatic
simplification arises in quantum mechanics: the inputs and outputs of
every admissible transformation will turn out to be a concatenation of
memory channels, and every admissible transformation will be itself
realized by a suitable concatenation of memory channels.  Notice that
in this approach memory channels are not assumed from the beginning,
but are derived on the basis of purely \emph{a priori} considerations
on the admissiblity of quantum maps.

\subsection{Quantum combs and admissible $N$-maps}\label{axi}

Quantum channels and operations are the most general transformations
of quantum states that satisfy the two minimal requirements of
linearity and complete positivity (see e.g. \cite{nielsen}).
Linearity is required by the probabilistic structure of quantum
mechanics.  Indeed, if we apply the transformation $\map C$ to the
state $\rho = \sum_i p_i \rho_i$---corresponding to a random choice of
the states $\{\rho_i\}$ with probabilities $\{p_i\}$---then the output
state must be a random choice of the states $\{\map C (\rho_i)\}$ with
the same probabilities, i.e. $\map C (\rho) = \sum_i p_i \map C
(\rho_i)$. For the same reason, we should also have $ \map C(p \rho) =
p \map C(\rho)$ for any $0\le p \le 1$.  These two conditions together
imply that $\map C$ can be extended without loss of generality to a
linear map on $\Lin {\sH_S}$, $\sH_S$ being the system's Hilbert
space. On the other hand, complete positivity is required if we want
the transformation $\map C$ to produce a legitimate output $\map C
\otimes \map I_A (\rho_{SA})$ when acting locally on a bipartite input
state $\rho_{SA}$ on $\sH _S\otimes \sH_A$: in this case, this means
that we want the output $\map C \otimes \map I_A (\rho_{SA})$ to be a
positive matrix for any positive input $\rho_{SA}$. We now raise the
level from states to channels, and ask which are the admissible
transformations of channels. Again, the minimal requirements for an
admissible transformation will be linearity and complete positivity.
Linearity is motivated in the very same way as for transformations of
states. Likewise, complete positivity is needed to ensure that the
transformation can be applied locally on a bipartite channel. This
investigation has been carried out in Ref. \cite{supermaps}.

Let us consider maps $\tilde{\map S}$ from linear maps $\map
T:\Lin{\sH_1}\to\Lin{\sH_2}$ to linear maps $\map
T':\Lin{\sH_0}\to\Lin{\sH_3}$. We say that $\tilde{\map S}$ is
admissible if {\it i)} it is linear and if {\it ii)} it preserves complete
positivity, also when it is applied locally on a bipartite map $\map
R$. More explicitly, condition {\it ii}) requires that if $\map R$
from $\Lin{\sH_1}\otimes \Lin{\sH_A}$ to $\Lin{\sH_2}\otimes
\Lin{\sH_B}$ is CP, then also $\map R'=(\tilde{\map
  S}\otimes\tilde{\map I})(\map R)$ from $\Lin{\sH_0}\otimes
\Lin{\sH_A}$ to $\Lin{\sH_3}\otimes \Lin{\sH_B}$ is CP. The
admissibility properties can be mathematically characterized if we
consider the conjugate map $\map S$ of $\tilde{\map S}$, defined as
follows
\begin{equation}
  \map S:=\C\circ\tilde{\map S}\circ \C^{-1},
\end{equation}
that transforms the Choi-Jamio\l kowski operator $T$ of $\map T$ into the Choi-Jamio\l kowski
operator $T'$ of the map $\map T'$ (see Fig. \ref{commdia}). Linearity of $\tilde{\map S}$ is
equivalent to linearity of $\map S$, while the second property for $\tilde{\map S}$ is equivalent to
complete positivity of $\map S$. Since $\map S$ is in one-to-one correspondence with $\tilde{\map
  S}$, we associate the Choi-Jamio\l kowski operator $S$ of $\map S$ to both of them. In the present
Section we will systematically use the map $\map S$ instead of $\tilde{\map S}$ for simplicity,
however the whole construction that follows must be intended as dealing with transformations of
transformations rather than with transformations of operators, thus generating an infinite hierarchy
of higher--rank quantum maps.

\begin{figure}
\epsfig{file=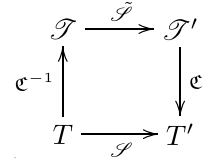,width=2.5cm}
  \caption{\label{commdia} The commutative diagram shows the relation
    between a transformation $\tilde{\map S}$ from linear maps $\map
    T$ to linear maps $\map T'$ and its conjugate $\map
    S:=\C\circ\tilde{\map S}\circ \C^{-1}$ through the Choi-Jamio\l kowski
    isomorphism, transforming Choi-Jamio\l kowski operators $T$ to Choi-Jamio\l kowski operators
    $T'$.}
\end{figure}


To tackle the characterization of all admissible quantum maps, we
start by defining a particular family of maps along with their Choi-Jamio\l kowski
operators:

{\definition \label{defcomb} A \emph{quantum $1$-comb} on
  $(\sH_0,\sH_1)$ is the Choi-Jamio\l kowski operator of a linear CP
  map from $\Lin{\sH_0}$ to $\Lin{\sH_1}$. For $N\ge 2$ a
  \emph{quantum $N$-comb} on $\tuplez{\sH}{N}$ is the Choi-Jamio\l
  kowski operator of an \emph{admissible $N$-map}, i.e. a linear
  completely positive map transforming $(N-1)$-combs on
  $\tupleu{\sH}{N-2}$ into 1-combs on $(\sH_{0},\sH_{2N-1})$.}

{\definition \label{defdetprob} A \emph{deterministic} 1-comb is the
  Choi-Jamio\l kowski operator of a channel. A \emph{deterministic}
  $N$-comb $S^{(N)}$ is the Choi-Jamio\l kowski operator of a
  \emph{deterministic} $N$-map, i.e. a map $\map S^{(N)}$ that
  transforms deterministic $(N-1)$-combs into deterministic 1-combs.
  A \emph{probabilistic} $N$-comb on $\tuplez{\sH}{N}$ is a positive
  operator $R^{(N)} \in \Lin{\bigotimes_{k=0}^{2N-1} \sH_k}$ such that
  $R^{(N)} \le S^{(N)}$ for some deterministic $N$-comb $S^{(N)}$ on
  $\tuplez{\sH}{N}$.  } \medskip

Definition \ref{defcomb} generates recursively an infinite family of
maps.  However, $N$-maps do not cover all possible maps one can define
in quantum mechanics.  Indeed, one might also consider maps from
$N$-combs to $M$-combs, take their Choi-Jamio\l kowski operators,
define maps thereof, and so on, with an exponential growth of the tree
of admissible quantum maps.  However, we will prove in subsection
\ref{nmmappes} that all admissible quantum maps can be reduced to
$N$-maps.

{\bf Remark (labeling of Hilbert spaces).} A quantum comb is defined
as an operator acting on an ordered sequence of Hilbert spaces.
Precisely, an $N$-comb is associated to an ordered sequence of $2N$
Hilbert spaces, which in Definition \ref{defcomb} are generically
labeled as $\sH_k$, $0\leq k\leq 2N-1$ (in the following we will need
to relabel spaces also with $1\leq k \leq 2N$); we will then denote
the set of deterministic $N$-combs on such $N$-tuple of spaces by
$\combz{\sH}{N}$ (or by $\combu{\sH}{N}$).

The assignment of labels can be easily done by exploiting an intuitive
diagrammatic representation of quantum combs.  In Fig.  \ref{diags} an
$N$-comb is denoted by a comb-like diagram with $N$ teeth labeled by
an ordered sequence of integers from left to right.  Quantum systems
are denoted by lines, and quantum operations (1-combs) by boxes.  Each
tooth $j$ ($0\leq j\leq N-1$) has an input (left) and output (right)
system, that, apart from cases that will be specified, are canonically
labeled $2j$ and $2j+1$, respectively.
\begin{figure}[h]
  \epsfig{file=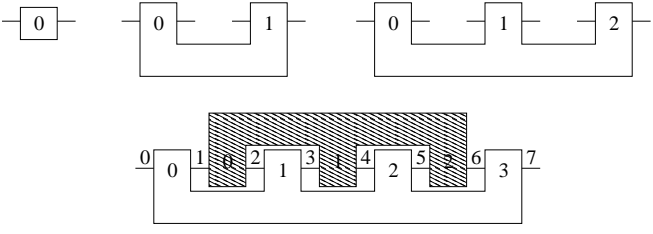,width=\columnwidth}
  \caption{\label{diags}In the first row we illustrate the
    diagrammatic representation of combs. A quantum system is
    represented by a line, a quantum operation (1-comb) by a box, a
    2-comb by a diagram with 2 teeth, and a 3-comb by a diagram with 3
    teeth. In the second row we represent the map corresponding to a
    4-comb, transforming the input 3-comb in an output 1-comb.}
\end{figure}
To describe the action of an $N$-comb on an $(N-1)$-comb, the Hilbert
spaces of the $N$-comb are labeled canonically as $\sH_k$, $0\leq k
\leq 2 N-1$, while the spaces of the input $(N-1)$-comb are labeled
as $\sH_k$, $1\leq k\leq 2N-2$. The output is an element of
$\operatorname{comb}(\sH_0,\sH_{2N-1})$, as in Definition
\ref{defcomb}.


In the following we characterize the convex set of quantum $N$-combs:

{\theorem \label{determin} A positive operator $S^{(N)}$ on
  $\bigotimes_{k=0}^{2N-1}\sH_k$ is a deterministic $N$-comb if
  and only if the following identity holds:
\begin{equation}
    \begin{split}
    &\Tr_{2j-1}[S^{(j)}]=I_{2j-2}\otimes S^{(j-1)},\quad 2\leq j\leq N\\
    &\Tr_1[S^{(1)}]=I_0,
  \end{split}
\label{normcondrecax}
\end{equation}
where $S^{(j)}, 1 \le j \le  N-1$ are deterministic $j$-combs.}


Before proving the theorem, we introduce two lemmas that will make the
proof simpler.

{\lemma \label{probabil} The set of positive operators $R^{(N)}$ such
  that $R^{(N)}\leq S^{(N)}$ for some $S^{(N)}$ satisfying
  Eq.~\eqref{normcondrecax} generates the positive cone in
  $\Lin{\bigotimes_{k=0}^{2N-1}\sH_k}$.}

\Proof The operator $J^{(N)}:=I/ (d_2 \dots d_{2k} \dots d_{2N})$,
where $d_k=\dim\sH_k$, clearly satisfies Eq.~\eqref{normcondrecax}. On
the other hand, any positive operator $T^{(N)}$ on
$\bigotimes_{k=1}^{2 N} \sH_k $, suitably rescaled, is smaller than
$J^{(N)}$, whence it is proportional by a positive factor to a
positive operator $R^{(N)}\leq J^{(N)}$.\qed

{\lemma \label{lemtwoprob} Consider two positive operators $R_i^{N}$,
  $i=1,2$, such that $R_i^{(N)}\leq S_i^{(N)}$ for some
  $S_i^{(N)}$ satisfying Eq.~\eqref{normcondrecax}. If
\begin{equation}\label{sametrace}
  \Tr_{2N-1}[R_1^{(N)}]=\Tr_{2N-1}[R_2^{(N)}],
\end{equation}
then there exists $T^{(N)}\ge0$ such that
$O_i^{(N)}:=R_i^{(N)}+T^{(N)}$ satisfy Eq.~\eqref{normcondrecax} for
$i=1,2$.}

\Proof Since $R_1^{(N)}\leq S_1^{(N)}$ there exists $T^{(N)}\ge0$ such
that $O_1^{(N)}:=S_1^{(N)} = R_1^{(N)} + T^{(N)}$. Due to Eqs.
(\ref{normcondrecax}) and (\ref{sametrace}) also the operator $O_2^{(
  N)}:= R_2^{( N)} + T^{(N)}$ satisfies Eq.~\eqref{normcondrecax}.\qed

{\bf Proof of theorem \ref{determin}.}  The proof proceeds by
induction. For $N=1$ the thesis is trivial: an operator $S^{(1)} \in
\Lin {\sH_0 \otimes \sH_1}$ is the Choi-Jamio\l kowski operator of a
channel from $\Lin{\sH_0}$ to $\Lin{\sH_1}$, if and only if $\Tr_1
[S^{(1)}] =I_0$ (see Eq.~\eqref{normchan}). We now suppose the theorem
holds for $1\leq M\leq N$, and show that it must hold also for $ N+1$.

{\bf Sufficient condition.} If $S^{(N+1)}$ is positive and satisfies
Eq.(\ref{normcondrecax}), then it is a deterministic $(N+1)$-comb.
Indeed, it is the Choi-Jamio\l kowski operator of the CP map $\map
S^{(N+1)}$, from $\Lin{\bigotimes_{k=1}^{2N} \sH_k}$ to
$\Lin{\sH_{2N+1} \otimes \sH_0}$, defined by
\begin{equation}
\begin{split}
  \map S&^{(N+1)} \left(R^{(N)} \right)\\
 &= \Tr_{2N, \dots, 1} \left[
    (I_{2N+1} \otimes R^{(N)T} \otimes I_0) S^{(N+1)}\right].
\end{split}
\end{equation} 
For deterministic $R^{(N)}$ the operator $\map S^{(N+1)}
\left(R^{(N)} \right)$ is the Choi-Jamio\l kowski operator of a
channel, because $\map S^{(N+1)} \left(R^{(N)} \right)\geq0$ and
\begin{align}
  \Tr&_{\sH_{2N+1}}[\map S^{(N+1)} \left(R^{(N)} \right)]=\nonumber\\
  &\Tr_{2N+1, \dots, 1} \left[ (I_{2N+1} \otimes R^{(N)T} \otimes I_0)
    S^{(N+1)}\right]=\nonumber\\
  &\Tr_{2N, \dots, 1} \left[ (R^{(N)T} \otimes I_0) (I_{2N}\otimes
    S^{(N)})\right]=\nonumber\\
  &\Tr_{2N-1, \dots, 1} \left[ (I_{2N-1} \otimes R^{(N-1)T} \otimes
    I_0) S^{(N)}\right]=I_0.
\end{align}
The final equality is obtained considering that by the induction
hypothesis $R^{(N-1)}$ is a deterministic $N-1$-comb, and by
hypothesis $S^{(N)}$ is a deterministic $N$-comb.

{\bf Necessary condition.}  Let $S^{(N+1)}$ be an $(N+1)$-comb and
$\map S^{(N+1)}$ be the corresponding map, which transforms a
deterministic $N$-comb $O^{(N)}\in\combu{\sH}{N}$ into a deterministic
$1$-comb $\map S^{(N+1)} (O^{(N)})\in\operatorname{comb}(\sH_0,\sH_{2N
  +1})$. Then, consider a couple of probabilistic $ N$-combs
$R_1^{(N)}, R_2^{(N)}$ on $\bigotimes_{k=1}^{2 N}\sH_k$, such that
\begin{equation}
  \Tr_{2  N}\left [R_1^{(N)} \right]=\Tr_{2  N}\left [R_2^{(  N)} \right].
\end{equation}
Since $R_i^{(N)}$ is probabilistic, by Def. \ref{defdetprob} there
exists a deterministic $N$-comb $Q_i^{(N)}$ such that $R_i^{(N)} \le
Q_i^{(N)}$. By lemma \ref{lemtwoprob} there exists $T^{(N)}$ such that
$O_i^{( N)} := R_i^{(N)} + T^{(N)}$ is deterministic for some $i=1,2$.
Then we have
\begin{equation}
\begin{split}
  \Tr_{2N+1}&\left[\map S^{(N+1)}\left(O_1^{(N)}\right)\right]=I_0\\
  &= \Tr_{2N+1}\left[\map S^{(N+1)} \left(O_2^{(N) }\right) \right],
\end{split}
\end{equation}
and consequently
\begin{equation}
  \Tr_{2N+1}\left[\map S^{(N+1)}\left(R_1^{(N)}\right)\right]=\Tr_{2N+1}\left[\map S^{(N+1)}\left(R_2^{(N) }\right)\right].
\end{equation}
In particular, by taking $R_2^{(  N)}= \sigma \otimes \Tr_{2   N}
[R_1^{(  N)}]$ for some state $\sigma$ on $\sH_{2   N}$, and using
Eq. (\ref{choicorrInv}) one has
\begin{align}
  \label{prima}  &\Tr_{2  N,\dots, 1}\left[(R_1^{(  N) T}\otimes I_{0})\Tr_{2  N+1}\left[S^{(  N+1)}\right ]\right]\\
\label{seconda} =&\Tr_{2  N,\dots, 1}\left[( R_1^{(  N) T} \otimes I_{0})
~ (I_{2   N} \otimes S^{(  N)})\right],
\end{align}
where we defined 
\begin{equation*}
  S^{(N)}:= \Tr_{2N+1,2N}[(I_{2N +1}
  \otimes \sigma^T \otimes I_{2  N-1} \otimes \cdots \otimes I_0)S^{(  N+1)}].
\end{equation*}
Comparing Eq. (\ref{prima}) with Eq. (\ref{seconda}), and using the
fact that probabilistic combs generate the cone of positive operators,
we then obtain
\begin{equation}
  \Tr_{2  N+1}[S^{( 
    N+1)}]=I_{2  N}\otimes S^{(  N)}.
\end{equation}
To conclude the proof, we need to prove that $S^{(N)}$ is a
deterministic $ N$-comb. To this purpose, define the CP map $\map
S^{(N)}$ from operators on $\bigotimes_{k=1}^{2N-2} \sH_k$ to
operators on $\sH_{0} \otimes \sH_{2N-1}$ as
\begin{equation}
\begin{split}
  \map S&^{(N)} (R^{(N-1)}) \\
  &:= \Tr_{2 N-2, \dots, 1} [ (I_{2 N-1} \otimes R^{( N-1) T} \otimes
  I_0) S^{( N)}].
\end{split}
\end{equation}
The map sends deterministic $  N-1$-combs in deterministic 1-combs.
Indeed, for any deterministic $  N-1$-comb $R^{(  N-1)}$ we have
\begin{align}
  \Tr&_{2   N-1} [\map S^{(  N)} ( R^{(  N-1)})] =\nonumber \\
  =&\Tr_{2   N-1, \dots ,1} [ (I_{2   N-1} \otimes R^{(  N-1)T} \otimes I_0 ) S^{(  N)}]\nonumber \\
  =& \Tr_{2 N + 1, \dots ,1} [ (I_{2 N+1} \otimes \sigma^T
  \otimes I_{2   N-1} \otimes R^{(  N-1)T} \otimes I_0)\nonumber \\
  &\quad S^{(  N+1)}]\nonumber \\
  =& \Tr_{2 N+1} [\map S^{( N+1)} (\sigma \otimes I_{2 N-1} \otimes
  R^{( N-1)} )]
\end{align}
for any state $\sigma$ on $\sH_{2 N}$.  Using the induction
hypothesis, we know that $\sigma \otimes I_{2 N-1} \otimes R^{( N-1)}$
is a deterministic $ N$-comb. Then the map $\map S^{( N)}$ is
deterministic, since
\begin{equation}
\begin{split}
  &\Tr_{2   N-1} [\map S^{(  N)} (R^{(  N-1)})] \\
  &= \Tr_{2 N+1} [\map S^{( N+1)} (\sigma \otimes I_{2 N-1}\otimes 
  R^{( N)})]= I_0.
\end{split}
\end{equation}  
This completes the proof. \qed

The deterministic $N$-combs $S^{(N)}\in\combz{\sH}{N}$ form a convex
set $\set{K}_N$ which is the intersection of the cone of positive
operators with the hyperplanes defined by Eq.  (\ref{normcondrecax}).
If we consider also the probabilistic combs, we have then the
following:

\medskip

{\bf Remark.} The cone generated by probabilistic $N$-combs in
  $\Lin{\bigotimes_{k=0}^{2 N -1} \sH_k}$ is the whole cone of
  positive operators.\label{probcomb}}
\medskip

Essentially, the above result implies that the only relevant cone in
quantum mechanics is the cone of positive operators. Another important
consequence of Theorem \ref{determin} is the isomorphism between
deterministic $N$-combs and Choi-Jamio\l kowski operators of $N$-partite channels
with memory: 


{\corollary \label{combmemch} A deterministic $N$-comb is also the
  Choi-Jamio\l kowski operator of an $N$-partite memory channel.}

\Proof Immediate from Theorems \ref{realchoi} and \ref{determin}. \qed


The following theorem finally proves that any deterministic map in the
hierarchy has a physical realization provided by a quantum memory
channel. Notice that, as we mentioned at the beginning of the present
section, the realization theorem regards the maps $\tilde{\map
  S}^{(N)}$ acting on $N-1$-maps $\tilde{\map T}^{(N+1)}$.

\begin{figure}[h]
\epsfig{file=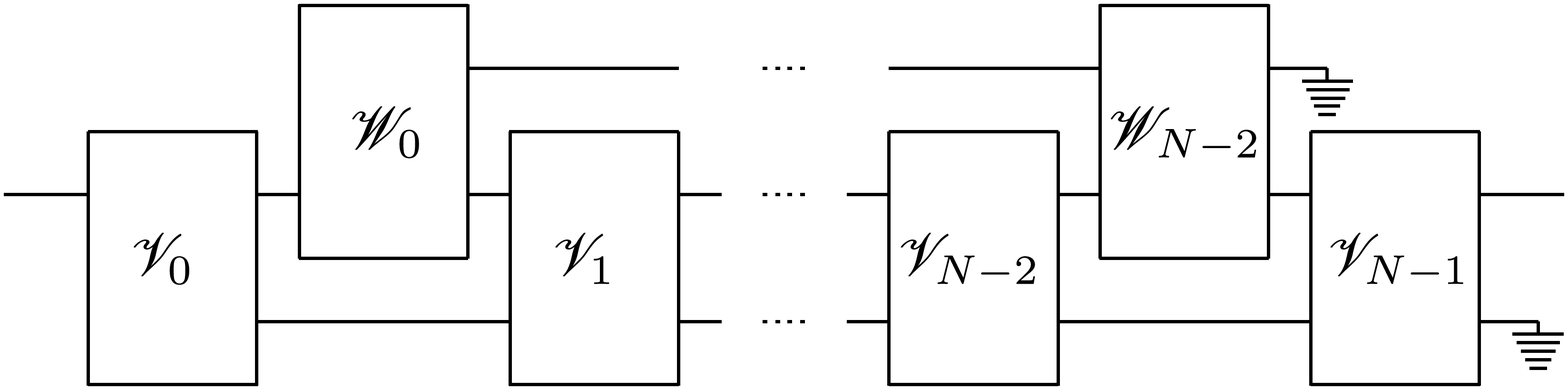,width=\columnwidth}
\caption{\label{phystwocombs} Realization of admissible $N$-maps by
  connection of memory channels. The input of the map is an
  $(N-1)$-comb, corresponding to a sequence of $N-1$ isometric
  channels $\{\map W_0, \dots, \map W_{N-2}\} $. The Choi-Jamio\l
  kowski operator of the map is an $N$-comb, corresponding to a
  sequence of $N$ isometric channels $\{\map V_0, \dots, \map
  V_{N-1}\}$.  The output of the map is obtained by connecting the
  free wires of the two memory channels.}
\end{figure}

{\theorem[Realization of admissible $N$-maps] \label{realcombs} For
  all $N$, any deterministic $N$-map $\tilde{\map S}^{(N)}$ can be
  achieved by a physical scheme tallying with the memory channel
  corresponding to its deterministic $N$-comb $S^{(N)}$. Let
  $T^{(N-1)}$ be any $(N-1)$-comb in $\combu{\sH}{N-2}$. The
  transformation
\begin{equation}
  \tilde{\map S}^{(N)}: \tilde{\map T}^{(N-1)} \mapsto \tilde{\map T}'^{(1)}=\tilde{\map S}^{(N)} \left(\tilde{\map T}^{(N-1)}\right)
\end{equation}
can be achieved by connecting the two memory channels represented
by $S^{(N)}$ and $T^{(N-1)}$ as in Fig. \ref{phystwocombs}.}

\begin{figure}[h]
\epsfig{file=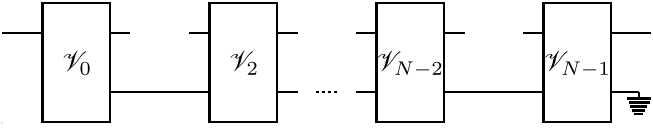,width=\columnwidth}
\caption{\label{memch} Identification of a quantum $N$-comb with the
  Choi-Jamio\l kowski operator of an $N$-partite memory channel. The teeth of the
  comb correspond to the isometries $\{\map V_0, \dots, \map V_{N-1}\}$ in
  the memory channel.}
\end{figure}

\Proof The statement is trivial for a deterministic 1-comb, which is a
quantum channel. Now, by induction, suppose that the transformation
$\tilde{\map T}^{(N-1)}$ corresponding to a deterministic $N-1$ comb
$T^{(N-1)}$ is realized by the $N-1$-partite memory channel having
Choi-Jamio\l kowski operator $T^{(N-1)}$, as in Fig. \ref{memch}. Let $W_0, i =1,
\dots, N-2$ be the Choi-Jamio\l kowski operators of the $n$ interactions occurring in
the memory channel, then $T^{(N-1)}$ can be expressed as
\begin{equation}
  T^{(N-1)}=\bar W_{N-2}*W_{N-1}*\dots*W_0,
\end{equation}
where the Choi-Jamio\l kowski operator $\bar X$ denotes the interaction described by
$X$ with the final ancilla traced out. By Corollary \ref{combmemch}
also $S^{(N)}$ is the Choi-Jamio\l kowski operator of a memory channel, then
$S^{(N)}$ can be expressed as
\begin{equation}
  S^{(N)}=\bar V_{N-1}*V_{N-2}*\dots*V_0,
\end{equation}
for suitable isometries $V_i$, where the link connects all the
spaces representing ancillae. The application of $\map
S^{(N)}=\C\circ\tilde{\map S}^{(N)}\circ \C^{-1}$ to
$T^{(N-1)}=\C(\tilde{\map T}^{(N-1)})$ provides
\begin{equation}
\begin{split}
  \map S^{(N)}&\left(T^{(N-1)} \right) =  S^{(N)}*T^{(N-1)}\\
&= \bar V_{N-1}*\bar W_{N-2}*V_{N-2}*\dots*W_0*V_0,
\end{split}\label{linkme}
\end{equation}
This proves that also the $N$-map $\tilde{\map S}^{(N)}$ can be
physically realized by a scheme as in Fig. \ref{memch}. Clearly,
Eq.~\eqref{linkme} prescribes that the action of $\tilde{\map
  S}^{(N)}$ on $\tilde{\map T}^{(N-1)}$ corresponds to connecting the
two memory channels associated to $S^{(N)}$ and $T^{(N-1)}$ as in Fig.
\ref{phystwocombs}.\qed

{\bf Remark (axiomatic approach to memory channels).}  It is worth
noticing that in the present setting $N$-partite memory channels are
derived from the recursive construction of admissible maps, rather
than being assumed as a particular type of channels with additional
causal structure. In this respect our approach differs with the
axiomatization put forward by Kretschamnn and Werner in Ref.
\cite{KW}, where memory channels are derived by starting from the
axiomatic definition of \emph{causal automata}, i.e. multipartite
quantum channels with the properties that \emph{i)} the output systems
at former times are not influenced by input systems at later times,
and \emph{ii)} the action of the channel is invariant under time
translations.  In the present approach, instead, the quantum memory
channel emerges in the Russian-dolls construction of maps on maps and
the causal structure is generated by the map-recursion. \par

In this respect, we would like to stress the interpretation of
Eq.~\eqref{normcondrec} as the mathematical translation of causal
ordering. In technical terms, this equation reflects the semicausality
property \cite{egge} for transformations occurring at teeth $j$ and
$i$, with $j<i$. This property is the mathematical translation of
independence of the $j$-th transformation from the $i$-th
transformation for $j<i$, namely the fact that information can be
transmitted from systems $j$ to system $i>j$, while the converse is
impossible.

\subsection{Tensor product combs and separable combs}

\begin{figure}[h]
  \begin{center}
    \epsfig{file=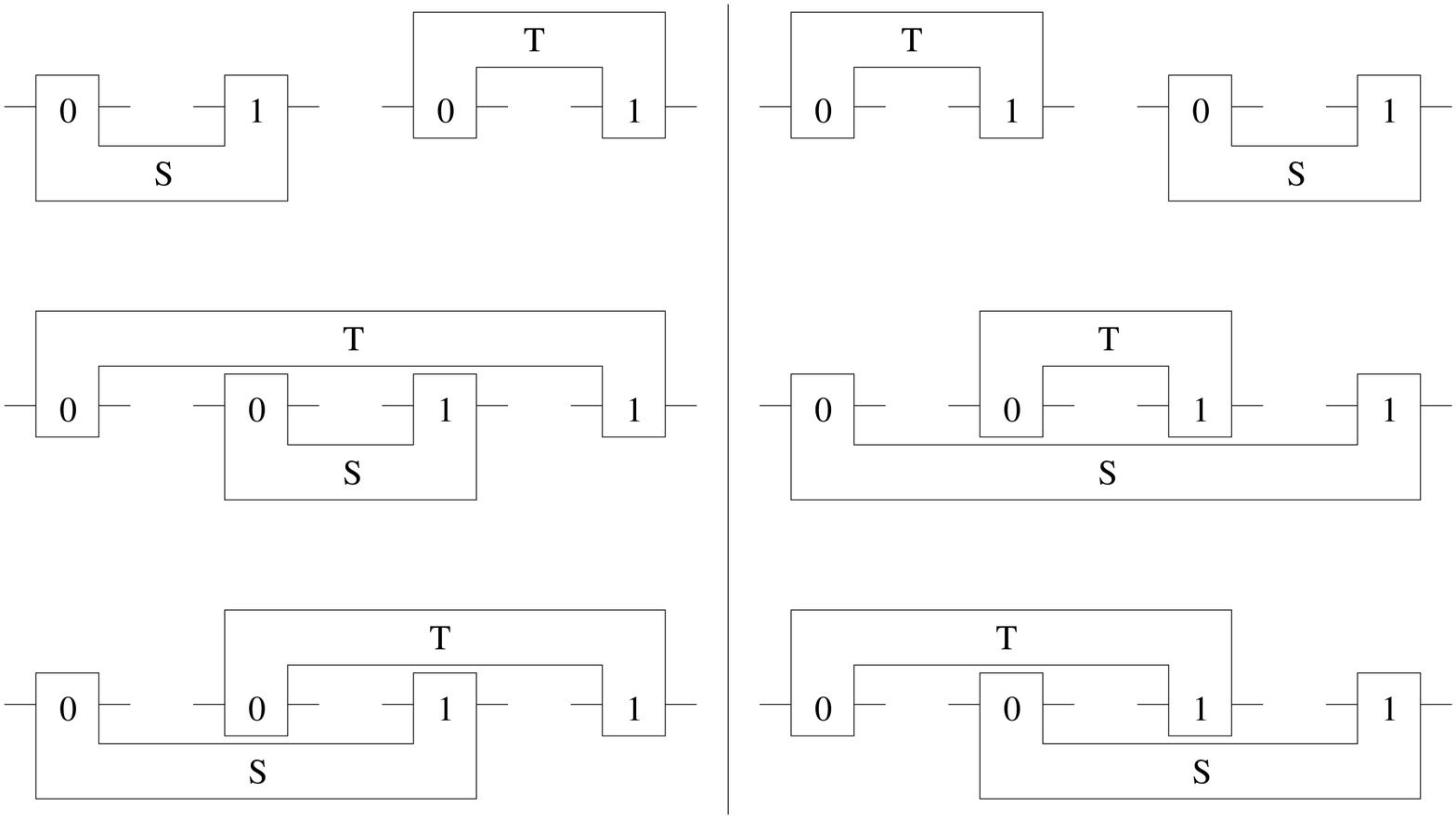,width=\columnwidth}\\
    (a)\\
    \epsfig{file=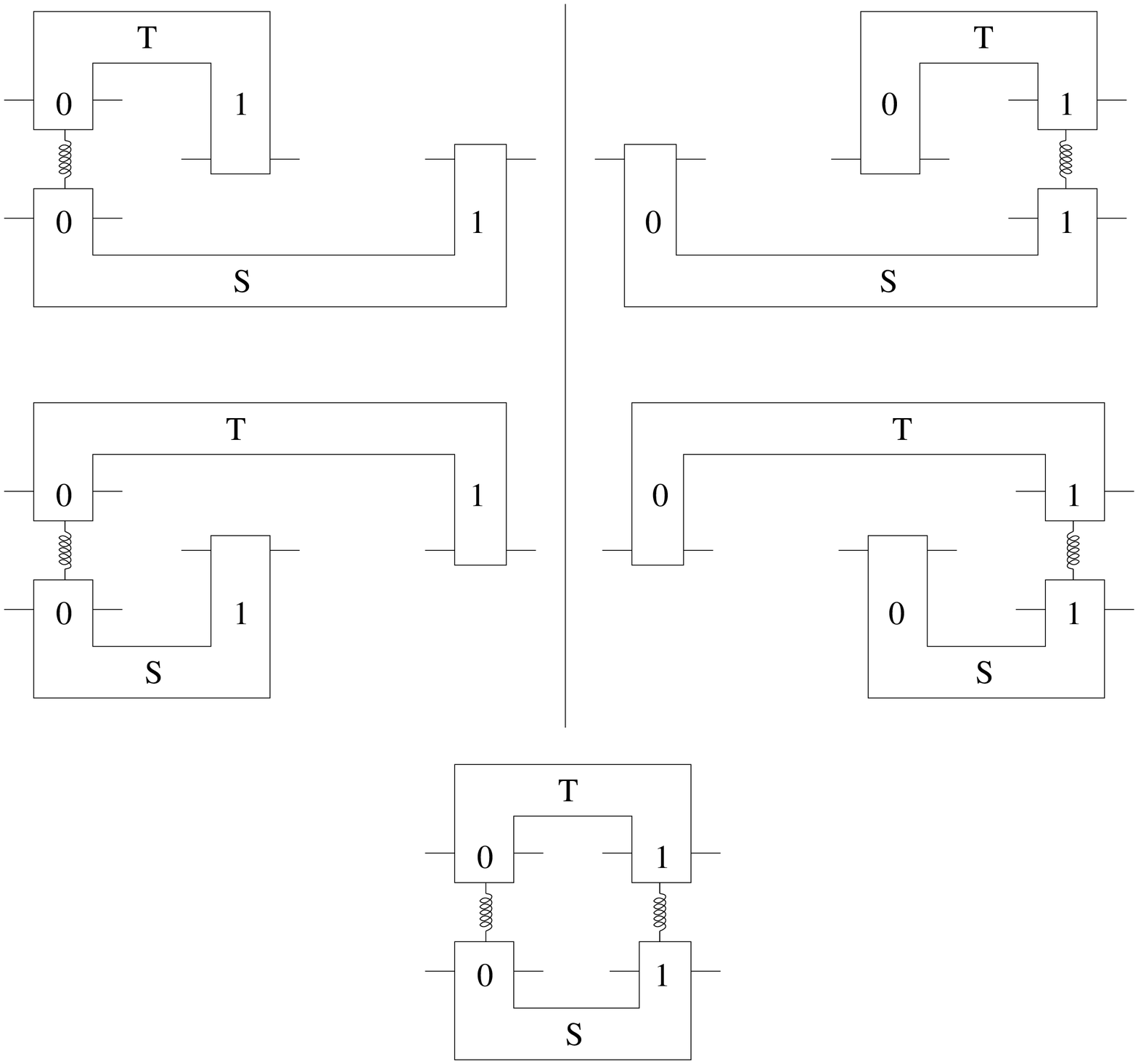,width=\columnwidth}\\
    (b)
  \end{center}
  \caption{\label{orderings} Diagrammatic representation of all
    different quantum combs arising from the tensor product of a
    2-comb $S^{(2)}$ with a 2-comb $T^{(2)}$.}
\end{figure}

As defined in subsection \ref{axi}, a quantum $N$-comb is a positive
operator over a tensor product of Hilbert spaces labeled by elements
of a totally ordered set. We now show how to combine two combs, say
$S^{(N)} \in \combz{\sK}{N}$ and $T^{(M)}\in \combz{\sK'}{M}$, in such
a way to obtain a new comb whose teeth are the teeth of both $S^{(N)}$
and $T^{(M)}$, e.~g. putting them in series, or in parallel, or in any
other way as in Fig. \ref{orderings}a. This corresponds to take the
tensor product of the operators $S^{(N)}$ and $T^{(M)}$ and suitably
reorder the Hilbert spaces of their teeth.  Instead of counting the
swap operators corresponding to such reordering, we will explicitly
show how to construct the resulting comb space. We need to consider
also situations as in Fig.  \ref{orderings}b, where two teeth, one
from each comb, are identified in a single tooth. It follows that the
general rule for the tensor product of combs is the following.

Let $S^{(N)} \in \combz{\sK}{N}$ and $T^{(M)}\in \combz{\sK'}{M}$ be two
quantum combs. Consider the following procedure
\begin{enumerate}
\item{Merge the sets of teeth of both combs in a single ordered set,
    preserving the relative ordering of each subset.}
\item{Consider the set $\set C$ of all couples of neighboring teeth
    containing a tooth from each comb, and select a subset $\set
    S\subseteq\set C$ of pairwise disjoint couples, whose cardinality
    is denoted by $S:=|\set S|$.}
\item{\label{ident}Identify each couple in $\set S$ in a single tooth
    (namely, the final tooth has input space given by the tensor
    product of the input spaces of the teeth in the couple, and
    similarly for the output space).}
\end{enumerate}
As a result, we obtain an ordered sequence of Hilbert spaces
$\tuplez{\sH}{L}$, with $L:=(N+M-S)$, which are the input and output
spaces of the teeth defined and ordered trough the previous procedure.

{\definition[Tensor product combs]\label{gentens} A \emph{tensor
    product comb} of $S^{(N)}$ and $T^{(M)}$ is the element of
  $\combz{\sH}{L}$ corresponding to the operator $S^{(N)}\otimes
  T^{(M)}$ with $\tuplez{\sH}{L}$ defined through the previous
  procedure.}

As a consequence, the tensor product of $S^{(N)}$ and $T^{(M)}$ is not
unique, depending on the merging of teeth and on the choice of the set
$\set S$ of identified couples. As an example, in Fig. \ref{orderings}
we represent all possible tensor products of two combs in the case
$N=M=2$.

{\bf Remark.} We could have enclosed a more general situation in the
definition of the tensor product of two combs, as follows. After
dividing the couples in the set $\set S$ into two sets $\set E$ and
$\set I$, proceed as in step \ref{ident} of the procedure with $\set
S$ replaced by $\set E$. As regards the remaining couples in $\set I$,
consider them to be {\em independent}. In this way, the final
operator $S^{(N)}\otimes T^{(N)}$ is considered as an element of a
subset $S_{\set I} \subseteq \combz{\sH}{L}$, such that if one swaps
couples in $\set I$ the resulting operator is in $\combz{\sH'}{L}$
where $\tuplez{\sH'}{L}$ is the corresponding reordering of
$\tuplez{\sH}{L}$. If $(i,j)\in\set I$, then any comb in $S_{\set I}$
satisfies the following identities
\begin{equation}
\begin{split}
  &\Tr_{2j+1}[R^{(j+1)}]=I_{2j}\otimes R'^{(j)},\\
  &\Tr_{2i+1}[R^{(j+1)}]=I_{2i}\otimes R''^{(j)}.
\end{split}
\end{equation}
A very simple example of 2-comb in $S_{\set I}$ with $\set
I=\{(0,1)\}$ is the comb of any convex combination of tensor product
channels $R:=\sum_{i}p_i C^{(i)}_{32}\otimes D^{(i)}_{10}$, where
$p_i$ are probabilities, $\Tr_3[C^{(i)}]=I_2$ and $\Tr_1[D^{(i)}]=I_0$
for all $i$. Another important example of 2-comb in $S_{\set I}$ with
the same $\set I$ as in the previous case is the following.  Consider
two channels, with input spaces $\sH_0\otimes\sH_A$ and
$\sH_2\otimes\sH_B$, respectively. The output spaces are $\sH_1$ and
$\sH_3$, respectively. If the channels are applied to a fixed state
$|\Psi\kk_{AB}\in\sH_A\otimes\sH_B$, then the resulting bipartite
channel from $\sH_0\otimes\sH_2$ to $\sH_1\otimes\sH_3$ can be viewed
as a 2-comb in both $\operatorname{comb}(\sH_0,\sH_1,\sH_2,\sH_3)$ and
$\operatorname{comb}(\sH_2,\sH_3,\sH_0,\sH_1)$. One might think that
all combs in $S_\set I$, with ${\set I}=\{(0,1)\}$, are achievable by
two local channels---one from $\sH_0\otimes\sH_A$ to $\sH_1$ and one
from $\sH_2\otimes\sH_B$ to $\sH_3$---applied to a bipartite, possibly
entangled ancillary state on $\sH_A\otimes\sH_B$. However, there exist
counterexamples to this conjecture, introduced in Refs.
\cite{bgnp,pianihoro}. In particular, the explicit counterexample of
Ref. \cite{pianihoro} corresponds to the following comb in
$\operatorname{comb}(\sH_0,\sH_1,\sH_2,\sH_3)$ with
$\sH_i\simeq{\mathbb C}^2$, that for sake of simplicity we write as an
operator on $\sH_1\otimes\sH_3\otimes\sH_0\otimes\sH_2$,
\begin{equation}
  R:=\frac12|I\kk\bb I|_{13}\otimes (I-P)_{02}+\frac12|\sigma_x\kk\bb\sigma_x|_{13}\otimes P_{02},
\end{equation}
where $P=|1\>\<1|\otimes|1\>\<1|$. One can verify that $R\in S_{\set
  I}$. However, any conceivable scheme for achieving the corresponding
map requires at least one round of classical information, in addition
to a shared entangled state.

\subsection{Admissible $(N,M)$-maps and higher order quantum
  maps\label{nmmappes}}

In this paragraph we give a definition of admissible quantum maps
$\tilde{\map S}^{(N,M)}$ from $N$-maps $\tilde{\map T}^{(N)}$ to
$M$-maps $\tilde{\map T}'^{(M)}$, showing that the corresponding
Choi-Jamio\l kowski operators are quantum combs themselves. This will
allow us to prove that the whole hierarchy of admissible quantum maps
defined axiomatically can be realized in terms of quantum memory
channels.  While a reasonable definition of a map from $N$-maps to
$M$-maps might seem to require only linearity and complete positivity,
such a definition turns out to be inadequate. As we will see in the
following, a consistent definition of admissible map involves an
additional requirement, that is \emph{compatibility with remote
  connections}.  To introduce this requirement, and the correct
definition, we first start from the definition involving only
linearity and complete positivity, and show the need of this
additional property.  As in the previous subsection, we will focus
attention on the conjugate maps $\map S^{(N,M)}:=\C\circ\tilde{\map
  S}^{(N,M)}\circ \C^{-1}$, transforming $N$-combs into $M$-combs.

{\definition\label{nmmap} An \emph{$(N,M)$-map} $\map S^{(N,M)}$ is a
  linear completely positive map transforming $\combz{\sK}{N}$ into
  $\combz{\sK'}{M}$.}  

{\definition \label{detprobnmmap} An $(N,M)$-map $\map S^{(N,M)}$ is
  \emph{deterministic} if it sends deterministic $N$-combs to
  deterministic $M$-combs. An $(N,M)$-map $\map R^{(N,M)}$ is
  \emph{probabilistic} if its Choi-Jamio\l kowski operator $R^{(N,M)}$
  satisfies $R^{(N,M)} \le S^{(N,M)}$ with $S^{(N,M)}$ the
  Choi-Jamio\l kowski operator of some deterministic map $\map
  S^{(N,M)}$.}

We have then the following equivalence:

{\lemma Let $\map S^{(N,M)}$ be a deterministic $(N,M)$-map.  Then
  $\map S^{(N,M)}$ is in one-to-one correspondence with a CP-map $\map
  S^{(N\times M)}$ that transforms tensor product operators $R^{(N)}
  \otimes O^{(M-1)}$ of deterministic $N$- and $(M-1)$-combs into
  deterministic 1-combs.}

Notice that the above statement does not involve tensor product combs,
but only tensor product operators: in other words, there is no fixed
total ordering of the Hilbert spaces on which the operator $R^{(N)} \otimes
O^{(M-1)}$ acts.

\Proof Suppose that $\map S^{(N,M)}$ maps an $N$-comb
$R^{(N)}\in\combz{\sK}{N}$ to $R'^{(M)} =\map S^{(N,M)}
(R^{(N)})\in\combz{\sK'}{M}$. In terms of Choi-Jamio\l kowski
operators, we have $R'^{(M)} = S^{(N,M)} * R^{(N)}$, where $S^{(N,M)}$
is the Choi-Jamio\l kowski operator of $\map S^{(N,M)}$. By
definition, the output comb $R'^{(M)}$ will be in turn the
Choi-Jamio\l kowski operator of a map ${\map R}'^{(M)}$ that
transforms $\combu{\sK'}{M-2}$ into
$\operatorname{comb}(\sK'_0,\sK_{2M-1}')$ as follows:
\begin{align}
  \map R'^{(M)} ( O^{(M-1)}) &= R'^{(M)} * O^{(M-1)}\nonumber\\
  &= S^{(N,M)} * R^{(N)} * O^{(M-1)}\nonumber\\
  &= S^{(N,M)} * ( R^{(N)} \otimes O^{(M-1)}),
\end{align}
where the last equality exploits Eq.~\eqref{tensorLink}.  Therefore,
the map $\map S^{(N,M)}$ induces a map sending tensor product
operators $R^{(N)} \otimes O^{(M-1)}$ to 1-combs.
\begin{equation}
  \map S^{(N\times M)} (R^{(N)}\otimes O^{(M-1)}) := S^{(N,M)} * ( R^{(N)} \otimes O^{(M-1)}).
\end{equation}
Clearly, if $R^{(N)}$ and $O^{(M-1)}$ are deterministic then $\map
S^{(N\times M)} (R^{(N)} \otimes O^{(M-1)})$ is deterministic.
Viceversa, given a CP-map $\map S^{(N\times M)}$ with Choi-Jamio\l
kowski operator $S^{(N\times M)}$, we can define the map $\map
S^{(N,M)}$ as $\map S^{(N,M)} (R^{(N)})= S^{(N\times M)} * R^{(N)}$.
If $\map S^{(N\times M)}$ sends products of deterministic combs into
channels, then $\map S^{(N,M)}$ is deterministic.\qed

The $(N,M)$-maps defined in Defs. \ref{nmmap} and \ref{detprobnmmap}
are then identified with maps that transform tensor products of $N$-
and $(M-1)$-combs into 1-combs. In other words, this means that if we
have at disposal a device implementing an $(N,M)$-map $\tilde{\map
  S}^{(N,M)}=\C^{-1}\circ\map S^{(N,M)}\circ \C$ from transformations
$\tilde{\map R}^{(N)}$ to transformations $\tilde{\map R}'^{(M)}$, we
can use it to transform a pair of independent transformations
$\tilde{\map R}^{(N)}\otimes\tilde{\map O}^{(M-1)}$ into a channel as
follows
\begin{equation}
  \tilde{\map S}^{(N\times M)}(\tilde{\map R}^{(N)}\otimes\tilde{\map O}^{(M-1)}):=[\tilde{\map S}^{(N,M)}(\tilde{\map R}^{(N)})](\tilde{\map O}^{(M-1)})
\end{equation}
However, we want to be able to use this device also locally on
transformations $\tilde{\map T}^{(N)}\otimes\tilde{\map T}'^{(M-1)}$
with multipartite input and output spaces, still producing a
legitimate output. If the map $\tilde{\map S}^{(N\times M)}$ can act
locally on two multipartite maps, the conjugate map $\map S^{(N\times
  M)}$ acts locally on the tensor product of two multipartite quantum
combs $T^{(N)}\otimes T'^{(M-1)}$. Since the physical implementation
of $T^{(N)}$ and $T'^{(M-1)}$ is provided by two memory channels, we
must also admit that the two input networks can be remotely connected
among themselves by some quantum memory.
Deciding which remote connections we assume to be possible is
equivalent to fixing a prescription for the causal ordering of the
Hilbert spaces in the tensor product, thus turning the tensor product
operator $R^{(N)} \otimes O^{(M-1)}$ into a tensor product comb, in
the sense of Definition \ref{gentens}. Moreover, the possibility of
remote connections entails the need of replacing the tensor product
comb $R^{(N)} \otimes O^{(M-1)}$---representing two independent
quantum networks---with a general $(N+M-S-1)$-comb
$R^{(N+M-S-1)}$---representing the compound network obtained by remote
connections. Therefore, in order for the map $\map S^{(N,M)}$ to
represent a legitimate deterministic quantum device, it should induce
a transformation of deterministic $(N+M-S-1)$-combs into channels.
In other words, $\map S^{(N\times M)}$ must be an admissible map on
$(N+M-S-1)$-combs defined through the tensor product.  This
crucial property, however, is not guaranteed by Definitions
\ref{nmmap} and \ref{detprobnmmap}.


As a consequence of the choice of one definition of tensor product,
the map $\map S^{(N\times M)}$ is then an admissible $N+M-S$-map $\map
S^{N+M-S}$ in the sense of Definition \ref{defcomb}, with respect to
the total ordering of the Hilbert spaces in the tensor product. The
above discussion motivates the following definition




{\definition[Admissible $(N,M)$-maps] \label{admnm} Let
  $(\sH_1,\dots\sH_{2L})$ be a reordering of spaces
  $(\sK_0,\dots,\sK_{2N-1})$ and $(\sK'_0,\dots,\sK'_{2M-3})$ as in
  definition \ref{gentens}, with $L=M+N-S-1$. An $(N,M)$-map $\map
  S^{(N,M)}$ from $\combz{\sK}{N}$ to $\combz{\sK'}{M}$ is
  \emph{admissible} if the associated map $\map S^{(L)}$ is an
  admissible $L$-map, sending $\combu{\sH}{L}$ to
  $\operatorname{comb}(\sH_0,\sH_{2L+1})$.}

{\definition \label{detprobadmnm} An admissible $(N,M)$-map is
  deterministic (probabilistic) if the corresponding map $\map
  S^{(N+M-S)}$ is deterministic (probabilistic).}

As an immediate consequence of the definition, we then have the
following identification: {\theorem  \label{unicombs} The Choi-Jamio\l kowski
  operator of an admissible $(N,M)$-map is a quantum $(N+M-S-1)$-comb. The
  comb is deterministic if and only if the map is deterministic.}

\Proof By definition the map $\map S^{(N,M)}$ has the same
Choi-Jamio\l kowski operator as $\map S^{(N+M-S-1)}$. The Choi-Jamio\l
kowski operator $S^{(N,M)}$ is then the Choi-Jamio\l kowski operator
of an admissible $(N+M-S)$-map, i.e. it is an $(N+M-S)$-comb. \qed

One might continue now the recursive generation of quantum maps by
defining admissible maps that transform admissible $(N,M)$-maps into
admissible $(K,L)$-maps. However, it is now clear that---as long as
independent teeth are excluded---such maps are in correspondence with
$N+M+K+L$-combs. Similarly, further levels in the hierarchy of
admissible maps are always admissible maps on combs, and hence combs
themselves. In other words, the whole hierarchy of admissible quantum
maps eventually collapses on $N$-maps $\tilde{\map S}^{(N)}$,
corresponding to quantum combs $S^{(N)}$. The conclusion of the whole
construction is the following property of universality of memory
channels, holding if one neglects the possibility of tensor product
combs with independent teeth

{\theorem{\bf (Universality of quantum memory channels)}
  \label{unimemo} The Choi-Jamio\l kowski operator of every
  deterministic admissible quantum map is a quantum comb $S^{(N+M)}$,
  and coincides with the Choi-Jamio\l kowski operator of a suitable
  sequence of memory channels. Any deterministic admissible quantum
  map is realized by interconnection of the input sequence of memory
  channels, corresponding to the input comb $T^{(N)}$, with the
  sequence corresponding to $S^{(N+M)}$.}  

\begin{figure}[h]
\epsfig{file=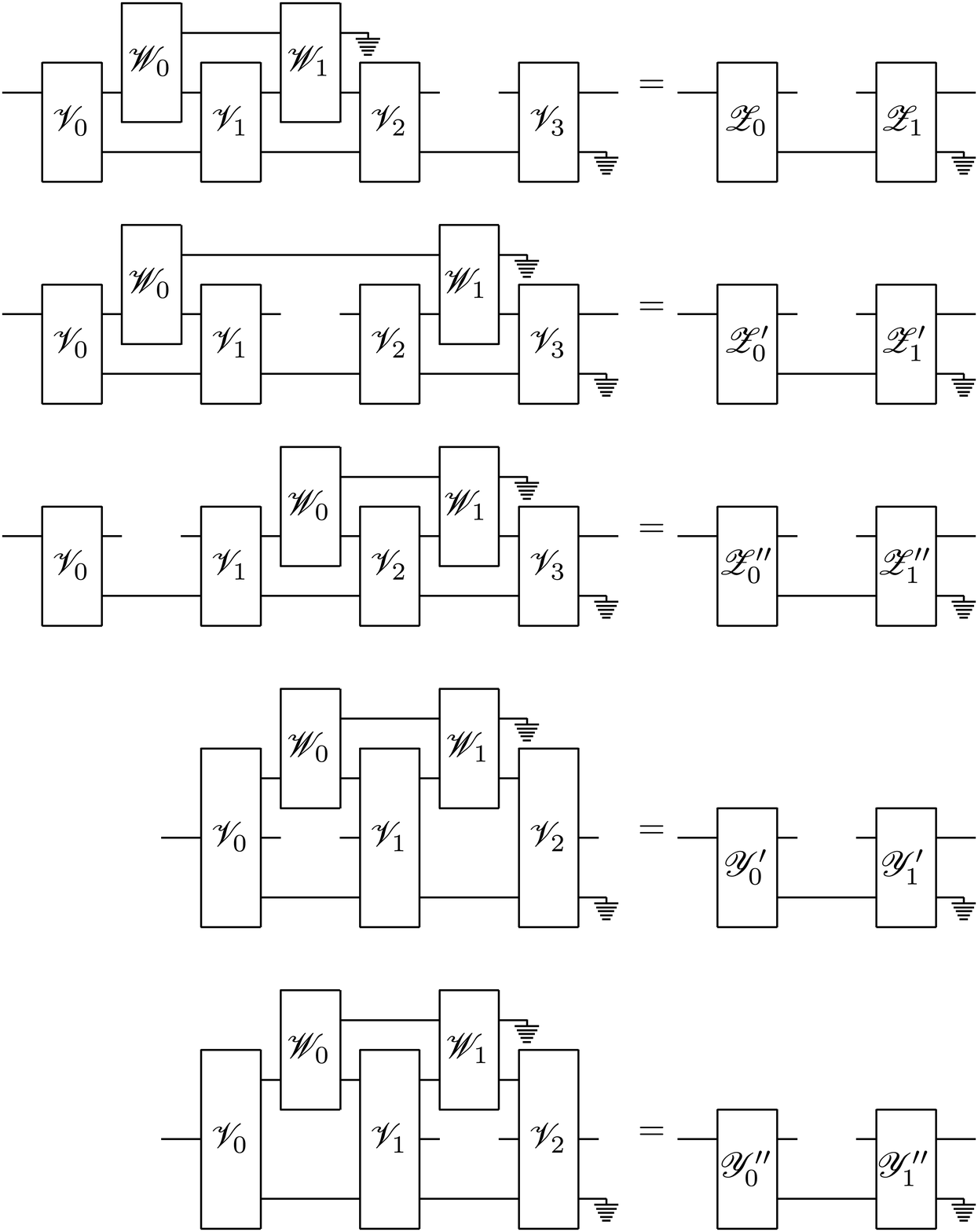,width=\columnwidth}
\caption{\label{fig:nmmap} The five possible realization schemes for
  admissible $(2,2)$ maps, that transform a 2-comb in a 2-comb.}
\end{figure}

\medskip
As an example, we show all possible schemes for admissible
$(2,2)$-maps in Fig.  \ref{fig:nmmap}.

{\bf Remark.} If we allow for independent teeth, the $(N,M)$-map as a
map on the tensor product $S^{(N)}\otimes T^{(M-1)}$ must be
admissible in a new sense, which is less restrictive than definition
\ref{admnm}. Indeed, it must only map the set $S_{\set I}$ to the set
of channels. Such maps are difficult to characterize and in general
they are not combs, as one can understand by the following example.
We will analyze the most elementary case, namely $S_{\set
  I}\subseteq\operatorname{comb}(\sH_1,\dots,\sH_4)$, with $\set
I=\{(0,1)\}$,
\begin{align}
  &\Tr_4[R]=I_3\otimes C_{21},\ \Tr_2[C]=I_1,\\
  &\Tr_2[R]=I_1\otimes D_{43},\ \Tr_4[D]=I_3.
\end{align}
These combs can be interpreted both as combs on
$\sH_4\otimes\sH_3\otimes\sH_2\otimes\sH_1$ and on
$\sH_2\otimes\sH_1\otimes\sH_4\otimes\sH_3$, and the set of most
general admissible maps transforming them into channels contains both
$3$-combs $A$ on
$\sH_5\otimes\sH_4\otimes\sH_3\otimes\sH_2\otimes\sH_1\otimes\sH_0$
and $B$ on
$\sH_5\otimes\sH_2\otimes\sH_1\otimes\sH_4\otimes\sH_3\otimes\sH_0$.
Thus, the most general admissible maps on $S_{\set I}$ include convex
combinations of the kind $C=pA+(1-p)B$, which are not combs in any
conceivable way, only satisfying $\Tr_{135}[C]=I_{024}$.  Moreover,
not all channels with the property $\Tr_{135}[C]=I_{024}$ do actually
represent admissible maps on $S_{\set I}$. We conclude the present
paragraph with the following open questions
\begin{enumerate}
\item{What is the most general realization scheme for elements of
    $S_{\set I}$?}
\item{How can we characterize and realize admissible maps on $S_{\set
      I}$?}
\end{enumerate}

\subsection{Generalized quantum instruments}

Here we consider an analogue of quantum instruments that is suitable to treat a generalized
measurement process where the measured object is a quantum network (described by its Choi-Jamio\l
kowski operator), rather than a quantum system (described by its state).  Such a generalized
instrument will associate to each measurement outcome the conditional Choi-Jamio\l kowski operator
of the quantum network. Notice that the number of input/output systems in the network can change in
this generalized measurement process, so that in principle we should consider probabilistic
transformations from networks with $N$ inputs/outputs (described by $N$-combs) to arbitrary networks
with $M$ inputs/outputs (described by $M$-combs). However, since we proved in the previous paragraph
that any admissible map from $N$-combs to $M$-combs is equivalent to an admissible map from
$(N+M-1)$-combs to $1$-combs, we can reduce without loss of generality the analysis of instruments
to this simpler case.  {\definition[Generalized $N$-instrument]. An $N$-instrument $\set I$ is a set
  of probabilistic $N$-combs $\{S^{(N)}_i~\}$ such that $\sum_i S^{(N)}_i$ is a deterministic
  $N$-comb.\label{definst}}

For simplicity we have confined here our attention to the case of
instruments with finite number of outcomes. The extension to the case
of measurements with arbitrary outcome space $\Omega$ is obtained by
defining the instrument as a \emph{Choi-Jamio\l kowski-operator
  valued measure} $S^{(N)} (B)$ \cite{covinst}, which associates to
any event $B \subseteq \Omega$ a probabilistic quantum $N$-comb
$S^{(N)} (B)$. The normalization of the measure amounts to the
requirement that $S^{(N)} (\Omega)$ is a deterministic $N$-comb.

{\theorem For any probabilistic $N$-comb $R^{(N)}$ there exists an
  $N$-instrument $\set I$ such that $R^{(N)}\in\set
  I$.\label{probininst}}

\Proof By definition, there exists a deterministic $N$-comb $S^{(N)}$
such that $S^{(N)}\geq R^{(N)}$. Then, $\tilde
R^{(N)}:=S^{(N)}-R^{(N)}\geq 0$ and $S^{(N)}\geq \tilde R^{(N)}$, then
$\tilde R^{(N)}$ is a probabilistic $N$-comb, and $\set
I:=\{R^{(N)},\tilde R^{(N)}\}$ is a generalized $N$-instrument\qed

A quantum $N$-instrument $\{S^{(N)}_i\}$ can be used to define a
family of probabilistic maps $\{\map S^{(N)}_i\}$ from $(N-1)$-combs
$R^{(N-1)}$ to Choi-Jamio\l kowski operators of quantum operations
$\{Q^{(1)}_i\}$, by means of
\begin{equation}
  Q^{(1)}_i = \map S^{(N)}_i (R^{(N-1)}) = S^{(N)}_i * R^{(N-1)}.
\end{equation}  
This means that the quantum network with Choi-Jamio\l kowski operator
$R^{(N-1)} $ is randomly transformed in one of the quantum operations
with Choi-Jamio\l kowski operators $\{Q^{(1)}_i\}$.  The realization
of any generalized instrument is given by the following

\begin{figure}
\epsfig{file=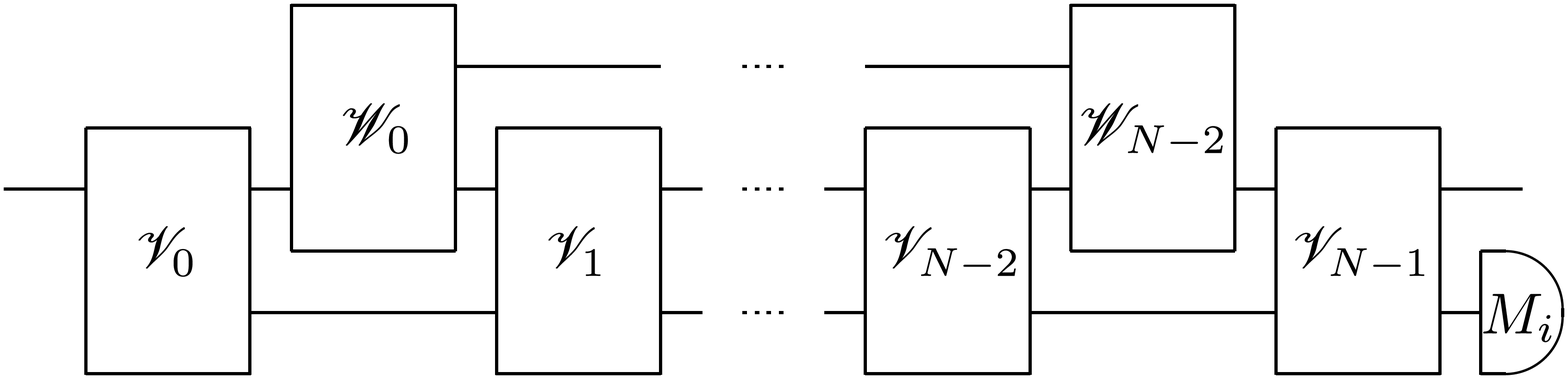,width=\columnwidth}
\caption{Realization of an $N$-instrument as a sequence of $N$
  isometric channels $\{\map V_0, \dots, \map V_{N-1}\}$ followed by a
  von Neumann measurement on the ancillary degrees of freedom. The
  input of the instrument is an $(N-1)$-comb, corresponding to a
  sequence of isometric channels $\{\map W_0, \dots, \map W_{N-2}\}$.
  Conditionally to outcome $i$, the output of the instrument is a
  quantum operation, which represents the input-output transformation
  of the whole composite network.  }\label{scheminst} 
\end{figure}

{\theorem[Realization of N-instruments]
  \label{realinst} Let $\set I=\{S^{(N)}_i~|~ i=1,\dots , k\}$ be an
  $N$-instrument, and let $R^{(N-1)}$ be an $(N-1)$-comb. The
  probabilistic transformations $\{\map S^{(N)}_i\}$ given by
  \begin{equation}
    \map
    S^{(N)}_i: R^{(N-1)} \mapsto \map S^{(N)}_i \left( R^{(N-1)} \right) = S^{(N)}_i
    * R^{(N-1)}
  \end{equation}
  can be achieved by a physical scheme as in Fig.  \ref{scheminst},
  involving isometric interactions of systems with quantum memories
  and a final von Neumann measurement on an ancilla with Hilbert space
  $\sH_A$ of dimension $\dim \sH_A = k$.}

\Proof Consequence of Theorems \ref{realcombs} and \ref{probnetchoi}.

 \subsection{Quantum testers and the generalized Born rule}

 Here we consider the particular case of admissible transformations of
 quantum networks in which input is a quantum $N$-comb and the output
 is just a probability. Such transformations are the analog of the
 customary POVMs describing measurements on quantum systems.

{\definition[Quantum tester]. An $N$-tester is a set of positive operators
  $\{P_i~|~ i = 1, \dots , k\}$ such that the quantities
  \begin{equation}\label{bornrule}
    p(i|R):=\Tr[P_i^T R]
  \end{equation}
  are probabilities for all deterministic $N$-combs $R$, i.e. $p(i|R) \ge 0$ and $\sum_i p(i|R)=1$.}

{\lemma \label{testinst} A set $\{P_i\}$ is an $N$-tester if and only if $\set I =\{ P_i\}$ is an
  $(N+1)$-instrument with $\dim\sH_0=\dim\sH_{2N+1}=1$.}

 \Proof The operator $T:=\sum_iP_i$ is a deterministic $(N+1)$-comb
 because it transforms any deterministic $N$-comb $R$ to the c-number
 1, which---regarded as a Choi-Jamio\l kowski operator---represents the only
 deterministic channel in a 1-dimensional Hilbert space\qed

 Since the tester is a particular case of generalized instrument, the
 normalization condition for the tester is given by
 Eq.~\eqref{normcondrec}, which in terms of $T$ becomes
\begin{equation}\label{normtest}
\begin{split}
  &T=I_{2N+2}\otimes\Theta^{(N)}\\
  &\Tr_{2j+1}[\Theta^{(j)}]=I_{2j}\otimes\Theta^{(j-1)},\ 1\leq j\leq n\\
  &\Tr_1[\Theta^{(0)}]=1
\end{split}
\end{equation}

The following corollary comes immediately from theorems \ref{realinst}
and \ref{testinst}.

{\theorem[Realization of testers] \label{cortes} Any $N$-tester
  $\{P_i~|~ i =1, \dots , k \}$ can be realized by an $(N+1)$-comb
  with $\dim\sH_0=1$ and $\dim \sH_{2 N+1} = k$ and by a von Neumann
  measurement on $\sH_{2N+1}$.}

\Proof The scheme is the same as in Theorem \ref{realinst}, except the
fact that the ancillary Hilbert space $\sH_A$ is now named
$\sH_{2N+1}$. Since the space $\sH_0$ is one-dimensional, the first
isometry $V_1$ is simply the preparation of an entangled state
$|\Psi\kk$. \qed

\begin{figure}[h]
\epsfig{file=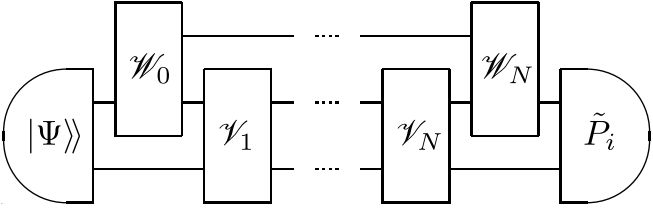,width=\columnwidth}
\caption{\label{schetest} Realization of a quantum $N$-tester as a
  probabilistic quantum network consisting of preparation of a pure input state
  $|\Psi\>\!\>$, isometric channels $\{\map V_1, \dots, \map V_N\}$, and a final
  measurement with POVM $\{\tilde P_i\}$.  The memory channel
  corresponding to the sequence of isometric channels $\{\map W_0, \dots, \map W_N\}$ is
  tested by connecting its wires with the wires of the tester and by
  running the resulting quantum circuit.}
\end{figure}

Notice that the probabilities $p(i|R)$ in the generalized Born rule
(\ref{bornrule}) arise as probabilities of outcomes in an experiment
as in Figure \ref{schetest}, where a pure entangled state $|\Psi\kk$
is prepared, and is evolved through a sequence of interactions until
the final measurement on $\sH_{2N+1}$.

We now provide an alternative proof for the realization of testers
that will be useful for later applications:

{\theorem[Realization scheme for testers] \label{realsqrt} Let
  $\{P_i\}$ be an $N$-tester with $\sum_i P_i = T $.  The tester can
  be split into a coherent part (state preparation and isometries) and a
  POVM, as in Fig.  \ref{schetest}.  The coherent part is described by
  a map $\map S$ sending $N$-combs to quantum states according to
  \begin{equation}
    \map S (R) = \sqrt T^T R \sqrt T^T.
  \end{equation}
  The POVM $\{\tilde P_i\}$ is given by
  \begin{equation}
    \tilde P_i = \sqrt {T^\ddag} P_i \sqrt{T^\ddag} + Q_i,
  \end{equation}
  where $T^\ddag$ is the Moore-Penrose generalized inverse of
  $T$---i.e.  $T^\ddag T = T T^\ddag = \Pi$, with $\Pi$ the projector
  on the support of $T$---and $\{Q_i\}$ is any set of positive
  operators such that $\sum_i Q_i = I - \Pi$.  The probabilities
  $p(i|R) = \Tr[ P_i R]$ are given by
  \begin{equation}
    p(i|R) = \Tr [\tilde P_i^T \map S (R)]. 
  \end{equation}}

\Proof Clearly, the set $\{ \sqrt{T^\ddag } P_i \sqrt{ T^\ddag} \}$ is
a POVM on the support of $T$, namely $\sum_i \sqrt {T^\ddag} P_i
\sqrt{T^\ddag} = \Pi$. One can consider a POVM $\{Q_i\}$ on the kernel
of $T$ with the same cardinality as $\{P_i\}$. It is now clear that
the operators
\begin{equation}
  \tilde P_i:=\sqrt{T^\ddag} P_i \sqrt{ T^\ddag}+Q_i
\end{equation}
define a POVM. Notice that by definition of generalized inverse one has $
  P_i =\sqrt T \tilde P_i \sqrt T$. 
The probabilities in the generalized Born rule Eq.~\eqref{bornrule}
are then obtained as follows
\begin{eqnarray}
  p(i|R)&=&\Tr[\sqrt T^T \tilde P_i^T \sqrt T^T R]\\
&=&\Tr[\tilde P_i^T \sqrt T^T R \sqrt T^T]\\
&=& \Tr[\tilde P_i^T \map S (R)].
\end{eqnarray}
Notice now that $\rho:=\sqrt T^T R\sqrt T^T$ is a state, since
$\Tr[\rho]=\Tr[T^T R]=1$ due to Eq. (\ref{normtest}). The map $\map S
(R) = \sqrt T^T R \sqrt T^T$ is clearly completely positive, and
transforms deterministic $(N-1)$-combs in states, which are Choi-Jamio\l kowski
operators of channels with one-dimensional input space $\sH_0$. Hence
$\map S$ is an $N$-comb and can be realized by a sequence of
isometries according to Theorem \ref{realcombs}. The first isometry is
necessarily a state preparation since $\sH_0$ is one-dimensional. \qed

The special case of $N$-testers with $N=1$, corresponding to
measurements on single channels, has been independently considered in
Ref.  \cite{ziman}, under the name process-POVM. In this case, the
realization scheme of the previous Theorem can be specialized to the
following

{\corollary[Realization scheme for 1-testers] Let $\{P_i\}$ be a
  1-tester and $C$ be the Choi-Jamio\l kowski operator of a channel. The
  normalization condition of the 1-tester is
  \begin{equation}
    \sum_i P_i=I\otimes \sigma,
  \end{equation}
  where $\sigma$ is a state. The probabilities $p(i|C) = \Tr [P_i^T C]$
  can be obtained by preparing a purification of the state $\sigma$,
  evolving it through the channel $\map C\otimes \map I$, and finally
  performing a measurement with POVM $\{\tilde P_i = (I \otimes \sqrt
  {\sigma^\ddag}) P_i (I \otimes \sqrt{\sigma^\ddag}) + Q_i\}$, with $Q_i$ as in
  Theorem \ref{realsqrt}.}

\Proof The normalization $T = I \otimes \sigma$ follows immediately
from Eq. (\ref{normtest}). According to Theorem \ref{realsqrt}, the
tester can be split in a coherent part and a POVM, with the coherent
part producing the state $\rho$ given by
\begin{equation}
\begin{split}
  \rho& =\map S (C)\\
  &= \sqrt T^T C \sqrt T^T\\
  &=(I\otimes \sqrt{ \sigma}^T )\map C\otimes\map I(|I\kk\bb I|)(I\otimes \sqrt{\sigma}^T)\\
  &=\map C\otimes\map I(|\sqrt{\sigma}\kk\bb\sqrt{\sigma}|).
\end{split}
\end{equation}
The last expression represents exactly the action of the channel $\map
C\otimes\map I$ on the purification $|\sqrt {\sigma}\kk$.\qed

In is worth noting the peculiarity of the case of 1-testers, where the
coherent part is simply achieved by preparing an entangled state on
which the variable channel $\map C$ is applied.  Typically, this is
not the case for $N>1$, as the general realization scheme given by
Fig. \ref{schetest} also contains the isometries $\{V_i~ i+1, \dots,
N-1\}$.  Such isometries generally play a crucial role, as they allow
to exploit memory effects that are extremely relevant when the
measured channel $\map C$ is an $N$-partite memory channel
\cite{memorydisc}.  As we will see in the following, $N$-testers are
the proper tool to treat the discrimination of two memory channels,
and to introduce a notion of distance between memory channels that is
related to statistical distinguishability.

\section{Application to discrimination and tomography of quantum networks}
\subsection{Distance and distinguishability}

According to the generalized Born rule (\ref{genborn}), two quantum
networks with the same quantum comb are experimentally
indistinguishable.  More generally, we are now in position to give a
notion of distance that captures the distinguishability of quantum
transformations.

Consider the problem of discriminating two $N$-partite memory
channels, described by the quantum $N$-combs $R_0$ and $R_1$,
respectively. In view of the discussion of the previous paragraphs,
this is enough to study the discrimination of all admissible
transformations in quantum mechanics. For simplicity, we discuss here
the problem of minimum error discrimination, in which the two memory
channels are given with prior probabilities $\pi_0$ and $\pi_1$.  Since
the most general transformation sending an $N$-comb in a set of
classical probabilities is given by an $N$-tester, any discrimination
experiment will be described by an $N$-tester $\{P_0, P_1 \}$ with
$P_0 +P_1 = T$ as in Eq.  (\ref{normtest}).  The average probability
of error is then given by
\begin{eqnarray}
  p_e &=& \pi_0 \Tr [P_1^T R_0 ] + \pi_1 \Tr[P_0^T R_1]\\
  &=& \pi_0 \Tr[ \tilde P_1^T \map S (R_0)] + \pi_1 \Tr [\tilde P_0^T \map S (R_1)]\\
  &=& \pi_0 -\Tr [(\pi_0
  \map S(R_0) - \pi_1 \map S (R_1)) \tilde P_0^T].
\end{eqnarray} 
where the map $\map S$ and the POVM $\{\tilde P_0, \tilde P_1\}$ are
as in Theorem \ref{realsqrt}.  The discrimination of the two memory
channels $R_0$ and $R_1$ is then reduced to the discrimination of the
states $\map S (R_0)$ and $\map S (R_1)$. Using Helstrom optimal
measurement \cite{helstrom} we get the bound
\begin{equation}
p_e \ge \frac{ 1 - \| \pi_0 \map S (R_0) - \pi_1 \map S (R_1) \|_1} 2,
\end{equation}
where $\| A \|_1 :=\Tr|A|$ is the usual trace-norm.  
Recalling that $\map S (R) = \sqrt T^T R \sqrt T^T$, and optimizing over $T$ finally gives following bound
\begin{equation}
p_e \ge \frac{ 1 - \max_{T}  \| \sqrt T^T ( \pi_0 R_0 - \pi_1 R_1) \sqrt T^T \|_1} 2,
\end{equation} 
where the maximum is taken over all operators $T\ge 0$ satisfying the
constraints in Eq. (\ref{normtest}). The bound is achievable, namely
once we have the optimal operator $T_{opt}$, and the Helstrom POVM
$\{\tilde P_0, \tilde P_1\}$ for the minimum error discrimination of
the states $\rho_i = \sqrt{T_{opt}}^T R_i \sqrt{T_{opt}}^T$ we can
define the optimal tester $\{P_i\}$ by $P_i= \sqrt{T_{opt}} \tilde P_i
\sqrt{T_{opt}}$. Theorem \ref{realsqrt} then ensures that there is a suitable scheme realizing the optimal tester.

The above discussion motivates the following definition:
{\definition[Distance between quantum combs] Let $R_0$ and $R_1$ be
  two $N$-combs. The distance between $R_0$ and $R_1$ is given by
\begin{equation}\label{opdist}
  d(R_0, R_1) = \frac12\max_{T} \left \|\sqrt T^T (R_0-R_1) \sqrt T^T \right \|_1,
\end{equation}
where $T$ is a positive semidefinite operator that satisfies Eq.
(\ref{normtest})}

This definition provides the suitable notion of distance between two
memory channels. This distance generalizes the notion of distance
based on the cb-norm \cite{paulsen} (alternatively called {\em diamond
  norm} \cite{kita}), that is typically used for quantum channels and
quantum operations. The cb-norm distance of two quantum
operations $\map O_0$ and $\map O_1$ from states on $\sH$ to states on $\sH$ is given by
\begin{equation}\label{cbdistance}
  d_{cb}( \map O_0 , \map O_1 ) = \frac12\sup_{n}  \sup_{\rho} \left \| [(\map O_0 -\map O_1) \otimes \map I_n] (\rho) \right \|_1, 
\end{equation}
where $\map I_n$ is the identity map on $\Lin {\mathbb C^n}$, and
$\sigma$ is a state on the extended Hilbert space $\sH \otimes \mathbb
C^n$. Using convexity of the trace distance and the finite
dimensionality of the input space $\sH$, the above expression can be
rewritten as \cite{max}
\begin{equation}\label{cb}
  d_{cb} (\map O_0,\map O_1) = \frac12\max_{\sigma} \left \| (I_{\sH} \otimes \sqrt\sigma^{T})\Delta(I_{\sH} \otimes \sqrt\sigma^{T}) \right\|_1,
\end{equation}
where $\sigma$ is a state on $\sH$, $\Delta:=O_0-O_1$, and
$O_0, O_1$ are the Choi-Jamio\l kowski operators of the quantum
operations $\map O_0, \map O_1$, respectively. Recalling that the
Choi-Jamio\l kowski operator of a quantum operation is a quantum comb
with $N=1$, and that for $N=1$ Eq. (\ref{normtest}) gives $T = I
\otimes \sigma$, we obtain
\begin{equation}
d_{cb} (\map O_0, \map O_1 ) = d ( O_0, O_1) \qquad {\rm for}~  N=1,
\end{equation}
namely for $N=1$ the cb-norm distance is a special case of distance between two
quantum combs.

Note that for $N$-partite memory channels $\map C_0$ and $\map C_1$
with Choi-Jamio\l kowski operators $C_0$ and $C_1$, respectively, the operational
distance introduced here is typically larger than the cb-norm
distance, i.e.
\begin{equation}
d (C_0, C_1) \ge d_{cb} (\map C_0, \map C_1).  
\end{equation} 
Indeed, Eq. (\ref{opdist}) involves maximization over all operators $T
\ge 0$ satisfying the constraints (\ref{normtest}), while Eq.
(\ref{cb}) involves maximization over operators of the special form
$T= I_{\sH} \otimes \sigma$, where now $\sH = \bigotimes_{k=0}^{N-1}
\sH_{2k}$ and $\sigma$ is a state on $\sH = \bigotimes_{k =0}^{N-1}
\sH_{2k+1}$.  The fact that for $nN1$ our distance can be strictly
larger than the cb-norm distance is due to the fact that the cb-norm
distance is related to discrimination in parallel schemes where the
unknown channel is applied to a large entangled state on $\sH^{\otimes
  2}$ and a collective measurement is finally performed on the
resulting state on $\sH \otimes \sH$, while the distinguishability of
two memory channels can be enhanced by using sequential schemes as in
Fig. \ref{schetest}.

\subsection{Informationally complete testers}

In the present section we introduce {\em informationally complete}
testers, namely testers $\{P_i\}$ such that the probabilities 
$p(i|R):=\Tr[P_i^TR]$ is sufficient to completely characterize the
(generally probabilistic) quantum comb $R$ on
$\bigotimes_{k=1}^{2n}\sH_k$. These testers are particularly important
for network tomography, in the very same way as informationally
complete POVMs describe possible tomographic experiments for quantum
states \cite{infoc}. Exploiting such testers in Ref. \cite{tomo}
tomography of quantum channels and operations has been optimized.
More precisely, the probabilities $p(i|R)$ is sufficient for the
reconstruction of $R$, if $p(i|R)$ allows to evaluate $\Tr[TR]$ for
all $T\in\Lin{\bigotimes_{k=1}^{2n}\sH_k}$ as follows
\begin{equation}
  \Tr[TR]=\sum_it_i\Tr[P_i^TR]=\sum_it_ip(i|R).
\label{infoc}
\end{equation}
From this condition the following definition comes straightforwardly.

{\definition[Informationally complete tester] The tester $\{P_i\}$ is
  informationally complete if and only if for all
  $T\in\Lin{\bigotimes_{k=1}^{2N}\sH_k}$ there exist coefficients
  $t_i$ such that $T=\sum_it_iP_i^T$.}

It is clear that this definition is an equivalent restatement of the
condition in Eq.~\eqref{infoc}. With the following theorem we prove
that informationally complete testers actually exist.

{\theorem For $\{\tilde P_i\}$ informationally complete POVM,  the tester with elements
  $P_i=\frac1{d_1\dots d_{2n-1}}\tilde P_i$ is informationally complete.\label{exist}}

\Proof If the POVM $\{\tilde P_i\}$ is informationally complete, then for all
operators $T\in\Lin{\bigotimes_{k=1}^{2n}\sH_k}$ one has
\begin{equation}
  T=\sum_it_i\tilde P_i^T.
\end{equation}
It is straightforward to verify that the coefficients $\tilde
t_i:=d_1\dots d_{2n-1}t_i$ expand $T$ on $P_i$. Moreover, $\{P_i\}$ is
a tester, since
\begin{equation}
  \sum_iP_i=\frac{I}{d_1\dots d_{2n-1}}, 
\end{equation}
which clearly satisfies the conditions in Eq.~\eqref{normtest}.\qed

In the following we will prove some theorems that will help
characterizing informationally complete testers.

{\theorem The operator $\Theta^{(N)}$ in Eq.~\eqref{normtest} providing
  the normalization of an informationally complete tester is
  invertible.}

\Proof Suppose that $\Theta^{(N)}$ is not invertible. Then the support of
$P_i$ is contained in the support of $I\otimes\Theta^{(N)}$. It is then
impossible that $\{P_i^T\}$ spans operators on the kernel of $I\otimes
\Theta^{(N)}$.\qed

{\theorem A tester $\{P_i\}$ is informationally complete iff it can be written as
  $P_i=(I\otimes\sqrt{\Theta^{(N)}})\tilde P_i(I\otimes\sqrt{\Theta^{(N)}})$,
  with $\{\tilde P_i\}$ informationally complete POVM and
  $\Theta^{(N)}$ invertible and satisfying identities \eqref{normtest}.}

\Proof Let us first suppose that $\{P_i\}$ is informationally
complete. Then $\sum_iP_i=I\otimes\Theta^{(N)}$ is invertible, and
since for all $T$ one has $T=\sum_i t_i P_i^T$ one also has
\begin{equation}
  \left(I\otimes\sqrt{\Theta^{(N)}}^T\right)T\left(I\otimes\sqrt{\Theta^{(N)}}^T\right)=\sum_i \tilde t_i P_i^T.
  \label{titild}
\end{equation}
If we now consider the POVM
$\tilde P_i:=(I\otimes\sqrt{\Theta^{(N)}}^{-1})P_i(I\otimes\sqrt{\Theta^{(N)}}^{-1})$,
we have clearly
\begin{equation}
  T=\sum_i \tilde t_i \tilde P_i^T,
\end{equation}
where the coefficients $\tilde t_i$ are the ones in
Eq.~\eqref{titild}. The set $\{P_i\}$ is then an informationally
complete POVM. On the other hand, if $\Theta^{(N)}$ is invertible and
satisfies Eq.~\eqref{normtest} and $\{\tilde P_i\}$ is an informationally
complete POVM, clearly
$\tilde P_i:=(I\otimes\sqrt{\Theta^{(N)}})P_i(I\otimes\sqrt{\Theta^{(N)}})$
is a tester. We can easily prove that it is informationally complete
by considering that since
$(I\otimes\sqrt{\Theta^{(N)-1}}^{T})T(I\otimes\sqrt{\Theta^{(N)-1}}^{T})=\sum_i
t_i P_i^T$ for all $T$, one has also $ T=\sum_it_i\tilde P_i^T$.\qed

In a completely analogous way, we can define informationally complete
testers for deterministic combs, which instead of separating the whole
$\Lin{\bigotimes_{k=1}^{2n}\sH_k}$ separate only the subspace $\sD$
spanned by deterministic combs. Notice that the set $\sD$ is given by
$\sD=\{X|\Tr_{2N-1}[X]=I\otimes Y\}$. The definition is then the
following

{\definition The tester $\{P_i\}$ is informationally complete for
  deterministic testers if and only if for all $T\in\sD$ there
  exist coefficients $t_i$ such that $T=\sum_it_iP_i^T$.}

Notice that this definition requires that the linear span of $\{P_i\}$
contains $\sD$ as a subspace. The existence theorem---analogous of
Theorem \ref{exist}---is trivial since any informationally complete
tester is also informationally complete for deterministic combs. On
the other hand, characterization theorems can be stated, but they are
beyond the scope of the present paper.

\section{Multiple-time states and measurements}

In this section we want to show that quantum combs and generalized instruments allow to treat in a
unified and simple framework the objects introduced in Ref. \cite{multitime} under the definitions
of {\em mltiple-time states} and {\em multiple-time measurements}.  Multiple time states correspond
to preparation of a state $|\Psi_0\>$ at time $t_0$ and subsequent post-selection by measurements
containing the Kraus operators $|\Psi_i\>\<\Phi_i|$ at times $t_i$, with $i=1,\dots,N-1$, and
finally post-selection by a {\em bra} $\<\Phi_N|$ at time $t_N$. The corresponding probabilistic
quantum comb is the following
\begin{equation}
\begin{split}
  &S=\bigotimes_{j=0}^N S_j,\\
  &S_N=|\Phi_N^*\>\<\Phi^*_N|,\ S_0=|\Psi_0\>\<\Psi_0|,\\
  &S_i=|\Psi_i\>\<\Psi_i|\otimes|\Phi_i^*\>\<\Phi^*_i|,\ 1\leq i\leq N-1.\\
\end{split}
\end{equation}
A multiple-time measurement is just a quantum operation with
multipartite Kraus operators $K^{(i)}_j$ for outcome $i$, such that
the probability of occurrence of the outcome $i$ for a multiple-time
state is provided by the expression
\begin{equation}
  p(i|S)=\frac{\sum_j\left|\<\Phi_1|\dots\<\Phi_{N}|K_j^{(i)}|\Psi_0\>\dots|\Psi_{N-1}\>\right|^2}{\sum_{lj}\left|\<\Phi_1|\dots\<\Phi_{N}|K_j^{(l)}|\Psi_0\>\dots|\Psi_{N-1}\>\right|^2}.
\end{equation}
In our formalism, a multiple-time measurement is described by a
generalized instrument $\{R_i\}$, with $R_i=\sum_j|K_j\>\< K_j|$, providing probabilities for
different outcomes on multiple-time states by the generalized Born
rule
\begin{equation}
  p(i|S)=\frac{\Tr[SR_i^T]}{\sum_j\Tr[SR_j^T]}.
\end{equation}
What the authors call an {\em history} is the outcome $i$ of the
generalized instrument. We want to stress that the approach to
multiple time states and measurements based on quantum combs provides
a simple answer to the following three fundamental questions left open by 
Ref. \cite{multitime}.
\begin{enumerate}
\item{{\em Given a Kraus operator, can we always find some multi-time
      measurement such that this operator represents a particular
      outcome of the measurement?} The answer is clearly yes, since by
    Corollary \ref{probcomb}, any positive operator, and in particular
    a rank one $|K\kk\bb K|$, suitably rescaled by a positive factor,
    provides a probabilistic comb, which in turn by Theorem
    \ref{probininst} can be included in a generalized instrument.}
\item{\em What are the conditions that a set of "histories" must
    satisfy in order to describe a measurement?} The answer is
  directly provided by the condition in Definition \ref{definst},
  representing the normalization of an admissible generalized
  instrument. More precisely, the Choi-Jamio\l kowski operators $R_i$
  of the histories must add to a deterministic comb, with the
  normalization conditions given by Eq. (\ref{normcondrec}).
\item{\em Is it possible that there are cases of sets of Kraus
    operators that do not lead to causality violations but still there
    is no actual way to implement them in quantum mechanics?} In this
  case the answer is negative. We provided indeed the causal
  interpretation of conditions \eqref{normcondrec}. Any multi-time
  measurement that does not violate causality satisfies the latter
  condition, and by Theorem \ref{realinst} this implies that the
  measurement is feasible as in Fig \ref{scheminst}.
\end{enumerate}

As an example we consider the same multi-time measurement as the
authors provide, implementing the measurement of the difference of the
values of the operator $\sigma_x$ at times $t_1$ and $t_2$ on a qubit,
$\sigma_x$ denoting the Pauli matrix. The measurement can be
summarized by the following generalized instrument
\begin{align}
  P_{+2}&=|-\>\<-|_{t_2}\otimes |-\>\<-|\otimes|+\>\<+|_{t_1}\otimes |+\>\<+|\nonumber\\
  P_{-2}&=|+\>\<+|_{t_2}\otimes |+\>\<+|\otimes|-\>\<-|_{t_1}\otimes |-\>\<-|\nonumber\\
  P_0&=|+\>\<+|_{t_2}\otimes|+\>\<+|\otimes|+\>\<+|_{t_1}\otimes |+\>\<+|\nonumber\\
  &+|-\>\<-|_{t_2}\otimes |-\>\<-|\otimes|-\>\<-|_{t_1}\otimes |-\>\<-|\nonumber\\
  &+|-\>\<+|_{t_2}\otimes |+\>\<-|\otimes|-\>\<+|_{t_1}\otimes |+\>\<-|\nonumber\\
  &+|+\>\<-|_{t_2}\otimes |-\>\<+|\otimes|+\>\<-|_{t_1}\otimes |-\>\<+|\nonumber\\
\end{align}
where $\sigma_x|\pm\>=\pm|\pm\>$. Clearly, the measurement outcomes
$\pm 2$ correspond to $\sigma_x(t_1)-\sigma_x(t_2)=\pm 2$, while $P_0$
corresponds to $\sigma_x(t_1)-\sigma_x(t_2)=0$.

\section{Other applications}

The general theory of quantum combs is useful for many applications, ranging in different branches
of quantum mechanics, like Quantum Information theory, quantum game theory and cryptography, quantum
metrology, and finally foundations of physics.\par

In Quantum Information combs provide an efficient and immediate description of networks, which is
the most suitable for optimization purposes. For example, quantum algorithms can be thought of as
testers on chains of unitaries, representing successive calls of quantum oracles. The optimization
of the tester for discrimination of oracle classes would provide the scaling of the performances of
the optimal algorithm with respect to the number of oracle calls, allowing for a definite
classification of the quantum complexity class for a wide class of problems. An example of
application in quantum information is optimal cloning of unitary gates, that was studied in Ref.
\cite{clonunit}, where combs were used to find the optimal physical device allowing to emulate two
uses of the same unknown unitary gate by actually running it only once.\par

In quantum game theory or quantum cryptography, quantum combs describe all conceivable
strategies/protocols of players/users. This has been already noticed in Ref. \cite{watrous} for
protocols in which only quantum systems are exchanged, without classical (\ie openly known)
communication.  The use of quantum combs provides a great simplification in the analysis of
cryptographic protocols, where one can use the operational definition of distance between strategies
of Eq.~\eqref{opdist} for search of equilibria and analysis of cheating startegies.  Moreover,
quantum combs provide the tool for the analysis of all those protocols that involve quantum and
classical communication in more than one direction, \eg for the evaluation of two-way channel
capacity. In order to include classical communication parallel to the quantum one needs to consider
sets of non superimposable orthogonal states, which can be easily taken into account using a
C$^*$-algebraic version of quantum combs, as it is done for channels \cite{QBCour}.

\par

In quantum estimation theory and quantum metrology, quantum combs provide the appropriate framework
for parameter estimation, since in the actual situation it is a unitary transformation that carries
the parameter to be estimated. In this case the old approach of Helstrom \cite{helstrom} and Holevo
\cite{holevo} optimizes the POVM for a given class of input states, then optimizes the state within
the class, and finally optimizes the class itself. Instead, the quantum tester provides the
optimization with a unified procedure, including the case of multiple uses, and even optimizing over
all possible dispositions of the uses. Moreover, as proved in Ref. \cite{memorydisc}, memory effects
turn out to be crucial in the discrimination of memory channels.

\par

Regarding the feasibility of quantum combs, all the possible implementations of qubits and their quantum
gates already largely explored for quantum computation are eligible also for the implementation of
quantum combs. A very promising scalable implementation of quantum combs is provided by optical
qubits in silicon waveguides \cite{obrien}.

Finally, we would like to mention one possible development of combs
for foundations of physics, in particular for the formulation of a
Quantum Theory of causally undetermined spacetime structures. This
suggestion comes from a striking analogy between quantum combs and a
quantum realization of the {\em causaloid} of Hardy
\cite{hardya,hardyb} a promising tool for the formulation of quantum
gravity.

\subsection{Admissible maps in general operational settings}

In sec. \ref{sec:axio} we proved that the whole hierarchy of linear transformations of any order in
Quantum Mechanics reduces to one level, corresponding to memory channels. Any admissible
transformation is physically achievable by a memory channel, namely a channel exploiting ancillary
systems as quantum memories that correlate successive uses. We proved this feature for the classical
and quantum combs. However, our proof exploits the detailed features of the theory, and it may not
hold more generally for any probabilistic theory \cite{BBLW07}. More precisely, we proved that: 1)
all admissible $N$-maps are realized by memory channels; 2) any admissible map (\ie $(K,L)$-map) is
indeed an $N$-map. One may wonder whether such features are generic for any probabilistic theory, or
if they are true only for the quantum-classical case.
\section{Conclusion}

In conclusion, we introduced a mathematical description of quantum networks in terms of Choi-Jamio\l
kowski operators, from two complementary points of view. The constructive approach is based on the
composition of Choi-Jamio\l kowski operators. Within this approach, it is possible to characterize
the properties of composite networks by the unified necessary and sufficient condition in Eq. (\ref{rec}).\par

The axiomatic approach starts from a completely different perspective, and defines admissible maps
on quantum objects in a recursive manner, starting from states and quantum operations and rising the
level to transformations of transformations, describing them through their Choi-Jamio\l kowski
operator. We proved that under minimal requirements such transformations correspond to memory
channels, and their admissibility implies feasibility.\par

All details of the theory of quantum networks are explored and thoroughly proved, including
properties of generalized instruments, testers and informationally complete testers, along with
discriminability criteria and operational distances between networks.

A comprehensive outline of applications and possible implementations of the theoretical objects
introduced in the paper is provided, including the description of multiple-time states and
multiple-time measurements. In particular, application of quantum combs to the description of
multi-time states and measurements shows the power of this approach, enabling us to answer three
important questions left open in Ref. \cite{multitime}.

Finally, we introduce the problem of classification of operational probabilistic theories in terms
of the structure of the hierarchy of admissible transformations, which could in principle elucidate
the peculiarity of Quantum Mechanics with respect to other theories.

\acknowledgments This work has been supported by EU FP7 program
through the STREP project CORNER.


\begin{thebibliography}{99}
\bibitem{saturated-index} Labelling Hilbert spaces in this way make us able to define the
  operator analog of the saturated-index pairs for multiplying tensors in general relativity. When
  multiplying only two operators only on some pairs of isomorphic tensor factors we will use the
  same name for the two Hilbert spaces, whereas we use different names for all other Hilbert spaces
\bibitem{cycles} Both restrictions are due to the convention that
  different uses of the same circuit are considered here as different
  circuits. Notice that the situation in which a quantum system
  undergoes the evolution $\map C^2$ by passing twice through the gate
  $\map C$ does not correspond to a loop, since the two successive
  uses of channel $\map C$ are represented as the cascade of two
  different channels.
\end{thebibliography}
\end{document}